\documentclass[sigconf]{acmart}

\usepackage{booktabs} 
\usepackage{graphicx}
\usepackage{epstopdf}
\usepackage{balance}  
\usepackage{amsfonts}
\usepackage{url}
\usepackage{subfigure}
\usepackage{algorithm}
\usepackage{amssymb,amsmath}
\usepackage{bbm}
\usepackage{multirow}

\usepackage{color}
\DeclareMathOperator*{\argmin}{arg\,min}

\definecolor{light-gray}{gray}{0.95}

\acmYear{2018}
\copyrightyear{2018}

\acmPrice{15.00}
\begin{document}
\title{Demystifying Core Ranking in Pinterest Image Search}
\author{Linhong Zhu}
\affiliation{%
  \institution{Pinterest \& USC/ISI}
}
\email{linhongz@acm.org}

\begin{abstract}
Pinterest Image Search Engine helps millions of users discover interesting content everyday. This motivates us to improve the image search quality by evolving our ranking techniques. In this work, we share how we practically design and deploy various ranking pipelines into Pinterest image search ecosystem. Specifically, we focus on introducing our novel research and study on three aspects: training data, user/image featurization and ranking models. Extensive offline and online studies compared the performance of different models and demonstrated the efficiency and effectiveness of our final launched ranking models.
\end{abstract}
\maketitle

\section{Introduction}\label{sec:intro}
Various researches on learning to rank~\cite{BurgesRankNet2005,burges2010ranknet,cao2007learning,chapelle2011yahoo,dehghani2017neural,joachims2002optimizing,geng2007feature,yin2016ranking,liu2009learning,zheng2008general} have been actively studied over the past decades to improve both the relevance of search results and the searchers' engagement. With the advances of learning to rank technologies, people might have a biased opinion that it is very straightforward to build a ranking component for the image search engine. This is true if we simply want to have a workable solution: in the early days of Pinterest Image Search, we built our first search system on top of Apache Lucene and solr~\cite{smiley2011apache,McCandlessLucene2010} (the open-source information retrieval system) and the results were simply ranked by the text relevance scores between queries and text description of images. 

However, in Pinterest image search engines, the items users search for are Pins where each of them contains a image, a hyperlink and descriptions, instead of web pages or on-line services. In addition, different user engagement mechanisms also make the Pinterest search process vary from the general web search engines.  We therefore have evolved our search ranking over past few years by adding various advancements that addressed the unique challenges in Pinterest Image Search. 

\begin{figure}[!t]
\centering
\subfigure[Pinteres users can perform various actions towards the results Pins of the query ``valentines day nails".]{\includegraphics[height=42.5mm]{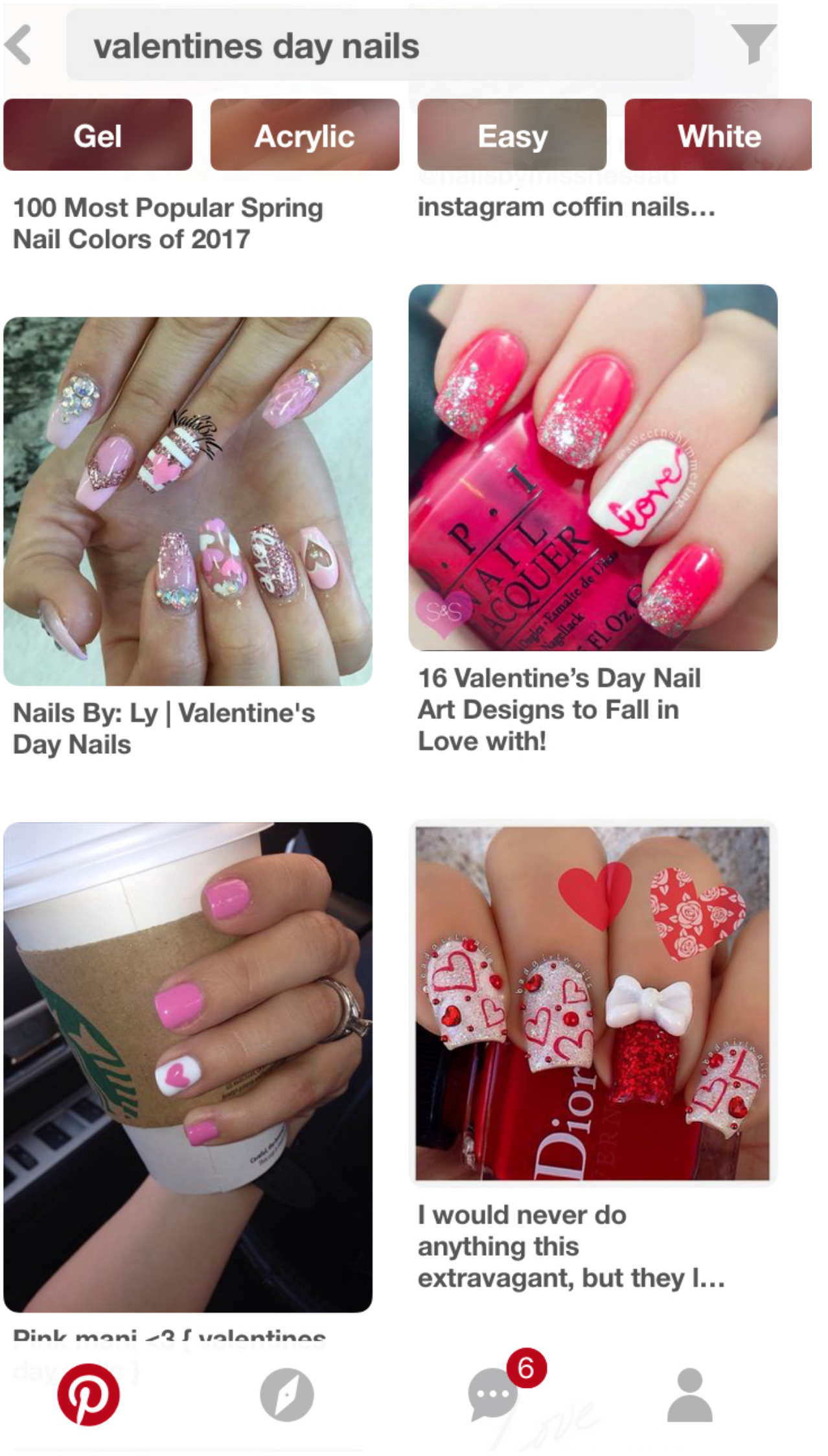}}
\hspace{0.1cm}
\subfigure[Close up: Click one pin leads to a zoom-in page. A further click on the ``save" button is called ``Repin''.]{\includegraphics[height=42.5mm]{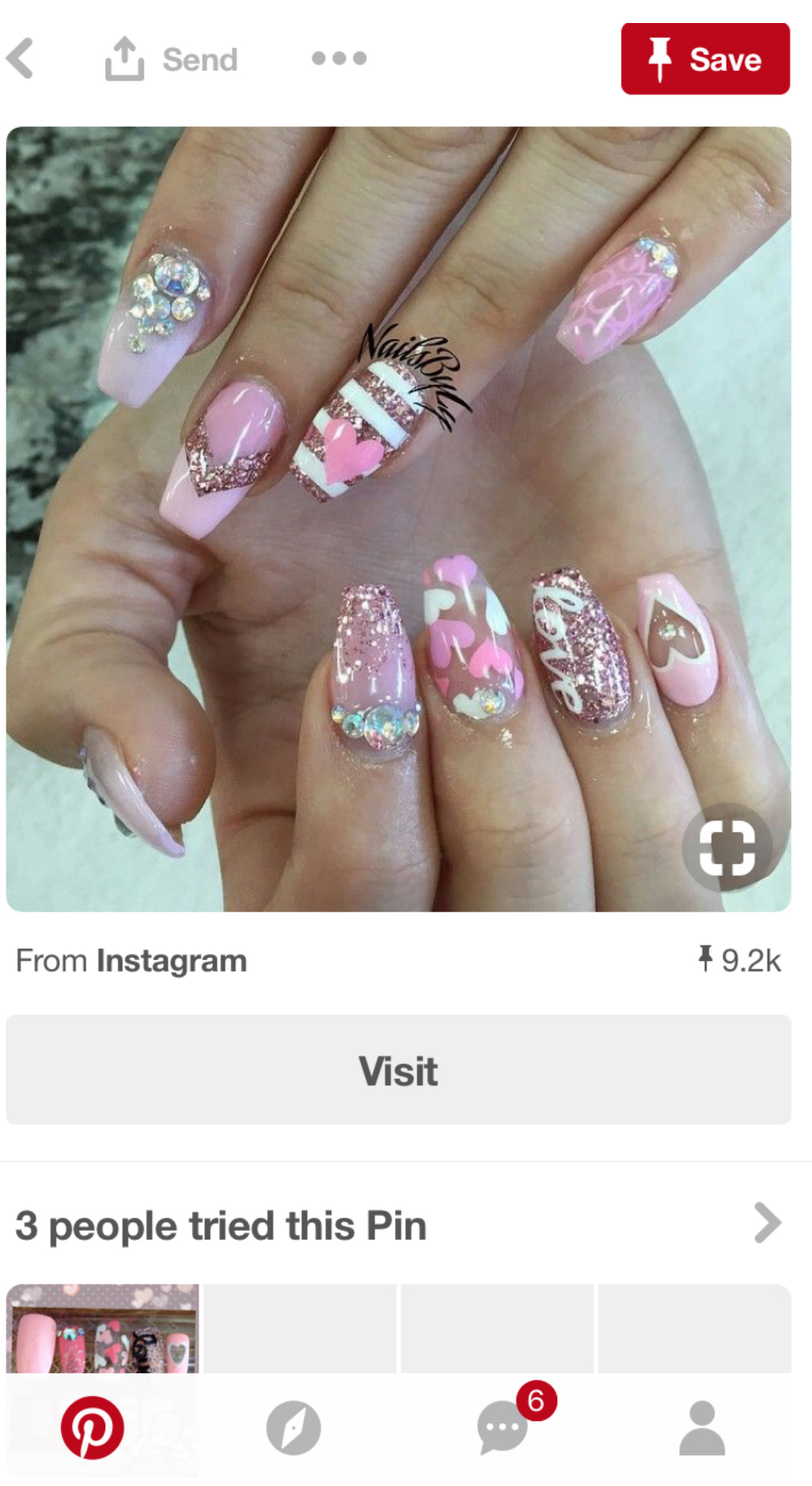}}
\hspace{0.1cm}
\subfigure[The second click on the close up page in (b) goes to the external website, is named as ``click'' in Pinterest.]{\includegraphics[height=42.5mm]{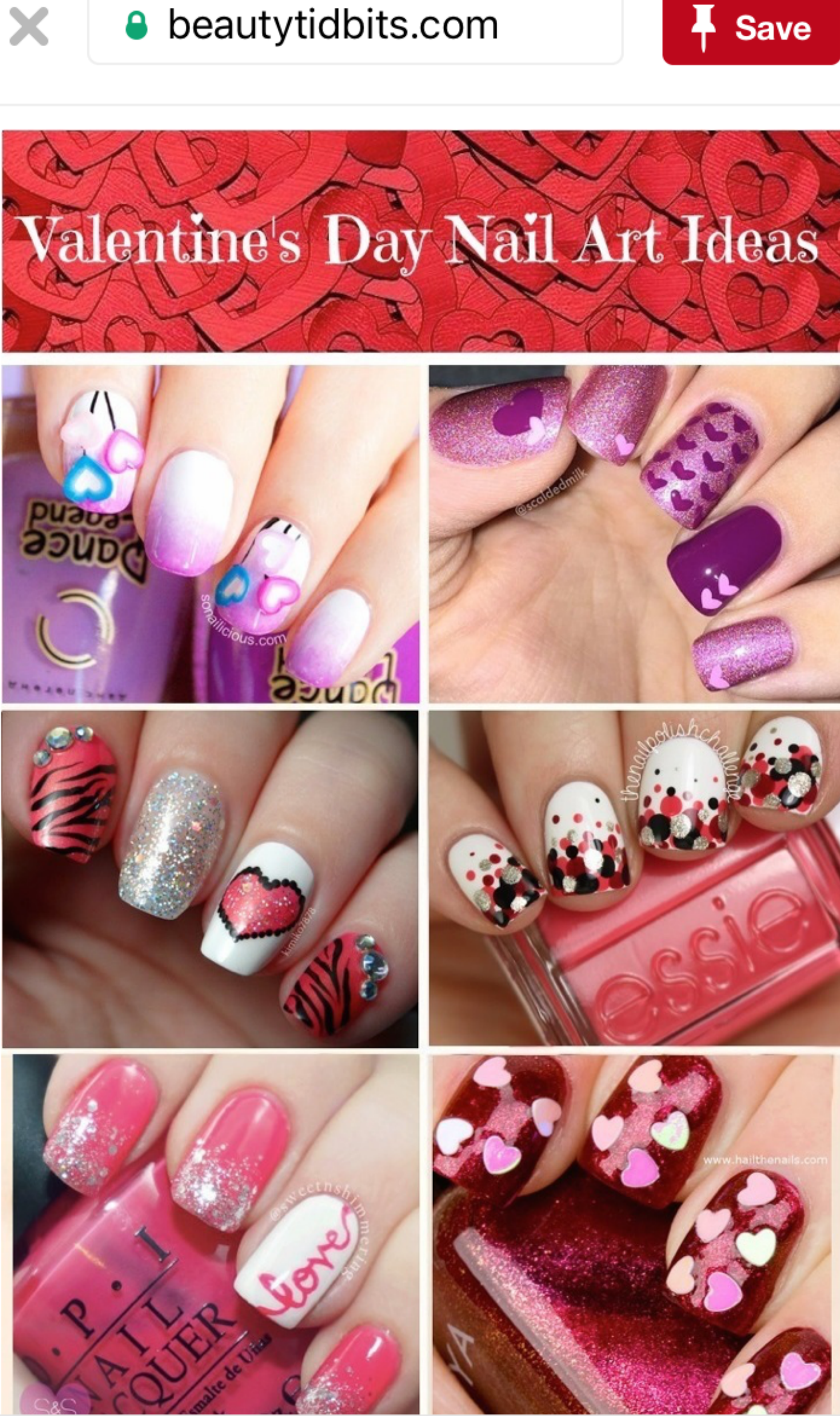}}
\vspace{-0.2cm}
\caption{Pinterest Image Search UI on Mobile Apps.}\label{fig:UI}
\vspace{-0.3cm}
\end{figure}

The first challenge rises from an important question: \emph{why users search images in Pinterest?} As shown in Figure~\ref{fig:UI}, Pinterest users (Pinners) can perform in total 60 actions towards the search results Pins such as ``repin'', ``click-through", ``close up'', ``try it" etc. In addition, users do have different intents while searching in Pinterest~\cite{lo2016understanding}: some users prefer to browse the pins to get inspirations while female users prefer to shop the look in Pinterest or search recipes to cook. On one hand, flexible engagement options help us to understand how users search for images and leverage those signals to provide a better ranking of search results; On another hand, the heterogeneity of engagement actions provides additional challenge about how we should incorporate those explicit feedbacks. In traditional search engine, a clicked result can be explicitly weighed more important than a non-clicked one; while in Pinterest ecosystem, it is very difficult to define a universal preference rule: is a ``try it" pin more preferable than a ``close up" pin, or vise versa? 

Another challenge lies in the nature of image items. Compared to the traditional documents or web pages, the text description of the image is much shorter and noisier. Meanwhile, although we understand that ``A picture is worth a thousand words", it is very difficult to extract reliable visual signals from the image.

Finally, much literature has been published on advanced learning to rank algorithms (see related work section) and their real-life applications in industry. Unfortunately, the best ranking algorithm to use for a given application domain is rarely known. Furthermore, image search engine system has much higher latency requirement than recommendation system such as News Feed, Friend Recommendation etc. Therefore, it is also very important to strike the balance between efficiency and effectiveness of ranking algorithms.

We thus address the aforementioned issues from three aspects:
\begin{description}
\item[Data] We propose a simple yet effective way to weighted combine the explicit feedbacks from user engagements into the ground truth labels of engagement training data. The engagement training data, together with human curated relevance judgment data, are fed into our core ranking component in parallel to learn different ranking functions. Finally, a model stacking is performed to combine the engagement-based ranking model with the relevance-based ranking model into the final ranking model.
\item[Featurization] In order to address the challenge in extracting reliable text and visual signal from pins, advancements in featurization that range from feature engineering, to word embedding and visual embedding, to visual relevance signal, to query intent understanding and user intent understanding etc. In order to better utilize the finding of why pinners use Pinterest to search images, extensive feature engineering and user studies were performed to incorporate explicit feedbacks via different types of engagement into the ranking features of both Pins and queries. Furthermore, the learned intent of users and other dozens of user-level features are utilized in our core machine learned ranking to provide a personalized image search experience for pinners.
\item[Modeling] We design a cascading core ranking component to achieve the trade-off between search latency and search quality. Our cascading core ranking filters the candidates from millions to thousands using a very lightweight ranking function and subsequently applied a much more powerful full ranking over thousands of pins to achieve a much better quality. For each stage of the cascading core ranking, we perform a detailed study on various ranking models and empirically analyze which model is ``better" than another by examining their performances in both query-level and user-level quality metrics.
\end{description}

The remainder of this work is organized as follows. In Section~\ref{sec:data}, we first introduce how we curated training data from our own search logs and human evaluation platform. The feature representation for users, queries and pins is presented in Section~\ref{subsec:feature}. We then introduce a set of ranking models that are experimented in different stages of the cascading ranking and how we ensemble models built from different data sources in Section~\ref{sec:ranking}. In Section~\ref{sec:expt}, we present our offline and online experimental study to evaluate the performance of our core ranking in production. Related work is discussed in Section~\ref{sec-related}. Finally we conclude this work and present future work in Section~\ref{sec:con}.

\section{Engagement and Relevance Data in Pinterest Search}\label{sec:data}
There are several ways to evaluate the quality of search results, including human relevance judgment and user behavioral metrics (e.g., click-through rate, repin rate, close-up rate, abandon rate etc). Therefore, a perfect search system is able to return both high relevant and high user-engaged results. We thus design and develop two relatively independent data generation pipeline: engagement data pipeline and human relevance judgment data pipeline. These two are seamlessly combined into the same learning to rank module. In the following, we share our practical tricks to obtain useful information from engagement and relevance data for learning module.
\subsection{Engagement Data}
Learning from user behavior was first proposed by Joachims~\cite{joachims2002optimizing}, who presented an empirical evaluation of interpreting click-through evidence. After that, click-through engagement Log has became the standard training data for learning to rank optimization in search engine. 
Engagement data in Pinterest search engines can be thought of as tuples $<q, u, (P, \mathcal{T})>$ consisting of the query $q$,  the user $u$, the set $P$ of pins the user engaged, and the engagement map $\mathcal{T}$ that records the raw engagement counts of each type of action over pins $P$. Note that here the notation user $u$ denotes not only a single user, but a group of users who share the same user feature representation. 

However, as introduced earlier in Figure~\ref{fig:UI}, when impression pins are displayed to users, they can perform multiple actions towards pins including click-through, repin, close-up, like, hide, comment, try-it, etc. While different types of actions provide us multiple feedback signals from users, they also bring up a new challenge: how we should simultaneously combine and optimize multiple feedbacks?

One possible solution is that we simply prepare multiple sources of engagement training data, each of which was fed into the ranking function to train a specific model optimizing a certain type of engagement action. For instance, we train a click-based ranking model, a repin-based ranking model, a closeup-based ranking model respectively. Finally, a calibration over multiple models is performed before serving the models to obtain the final display. Unfortunately, we tried and experimented with hundreds of methods for model ensemble and calibration and was unable to successfully obtain a high-quality ranking that does not sacrificing any engagement metric.

Thus, instead of calibrating over the models, we integrate multiple engagement signals over the data level. Let $l(p\mid q, u)$ denote the engagement-based quality label of pin $p$ to the user $u$ and query $q$. To shorten the notation, we simply use $l_p$ to denote $l(p\mid q, u)$ when the given query $q$ and user $u$ can be omitted with ambiguity. We thus generate the engagement-based quality label set $L$ of pins $P$ as follows. 

For each pin $p\in P$ with the same keyword query $q$ and user features $u$, the raw label $l_p$ is computed as a weighted aggregation of multiple types of actions over all the users with the same features. That is,
\begin{equation}\label{eq:action_sum}
l_p = \sum_{t\in \mathcal{T}}w_tc_t
\end{equation}
where $\mathcal{T}$ is the set of engagement actions, $c_t$ is the raw engagement count of action $t$ and $w_t$ is the weight of a specific action $t$. The weight of each type of action $w_t$ is reversely proportional to the volume of each type of action. 

We also normalize the raw label of each pin based on its position in the current ranking and its age to correct the position bias and freshness bias as follows:
\begin{equation}
l_p = l_p(\frac{1}{\log(\texttt{age}_p /\tau) + 1.0} + e^{\lambda\texttt{pos}_p})
\end{equation}
where $\texttt{age}_p$ and $\texttt{pos}_p$ are the age and position of pin $p$, $\tau$ is the normalized weight for the ages of pins, and $\lambda$ is the parameter that controls the position decay.

Another challenge in generating a good quality engagement training data is that we always have a huge stream of negative training samples but very few positive samples that received users' engagement actions. To avoid over learning from too many negative samples, two pruning strategies are applied:
\begin{enumerate}
\item Prune any query group $(q, u)$ and its training tuples $<q, u,$ $(P, \mathcal{T}, L)>$ that does not contain any positive training samples (i.e., $\forall p\in P$, $l_p\in L$, $l_p\leq 0$).
\item For each query group, randomly prune negative samples if the number of negative samples is great than a threshold $\delta$ (i.e., $|\{p\mid p \in P, l_p \leq 0\}| \leq \delta$).
\end{enumerate}

With the above simple yet effective ways, an engagement-based data can be automatically extracted from our Pinterest search Logs.
\subsection{Human Relevance Data}
\begin{figure}[!t]
\centering
\includegraphics[width=\columnwidth]{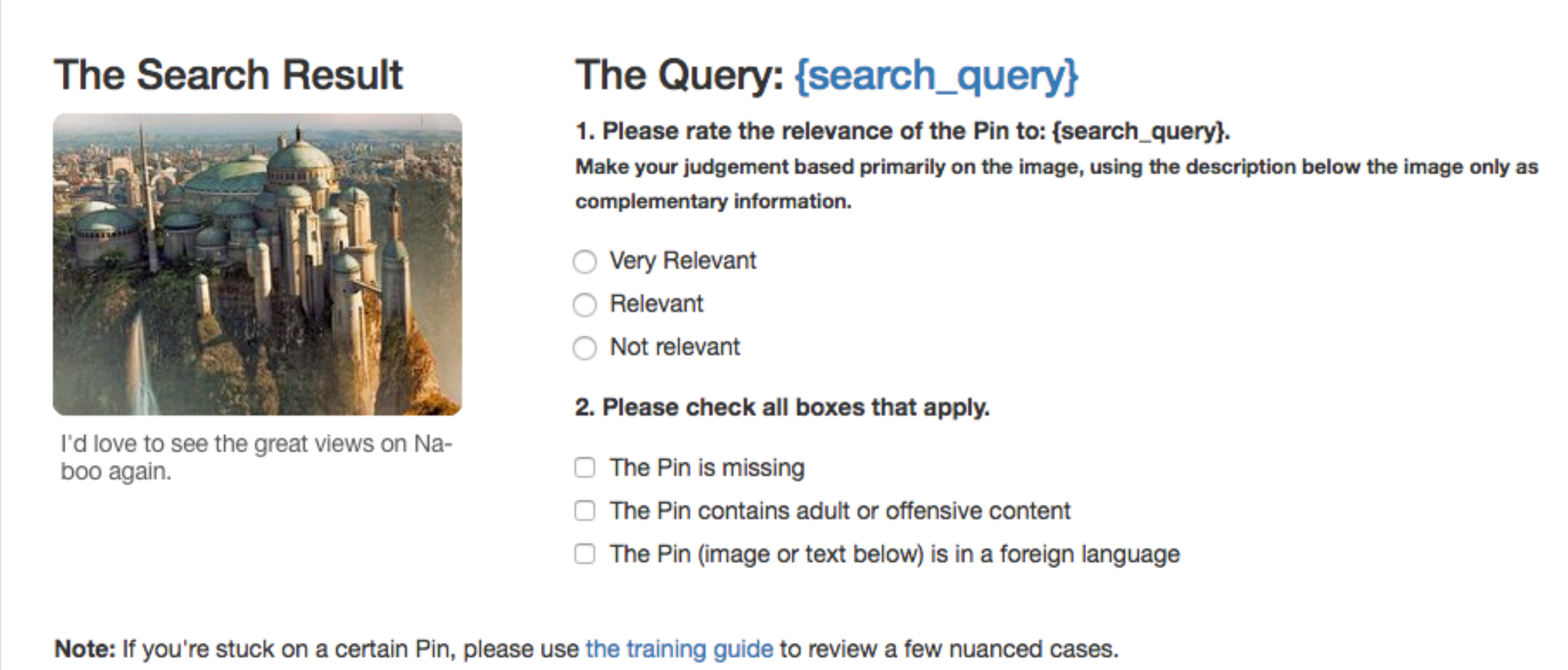}
\caption{Template for rating how relevant a pin is to a query.}\label{fig:template}
\vspace{-0.3cm}
\end{figure}

While the aggregation of large-scale unreliable user search session provides reliable engagement training data with implicit feedback, it also brings up the bias from the current ranking function. For instance, position bias is one of these. To correct the ranking bias, we also curate relevance judgment data from human experts with in-house crowd-sourcing platform. The template for rating how relevant a Pin is to a query is shown in Figure~\ref{fig:template}. Note that each human expert must be a core Pinterest user and pass the golden-set query quiz before she/he can start relevance judgment in a three-level scale: very relevant, relevant, not relevant. The raw quality label $l_p \in [0, 2]$ is thus averaged over ratings of all the human experts. 

\subsection{Combining Engagement with Relevance}
\begin{figure}[!t]
\centering
\subfigure[Distribution of engagement scores]{\includegraphics[width=0.495\columnwidth]{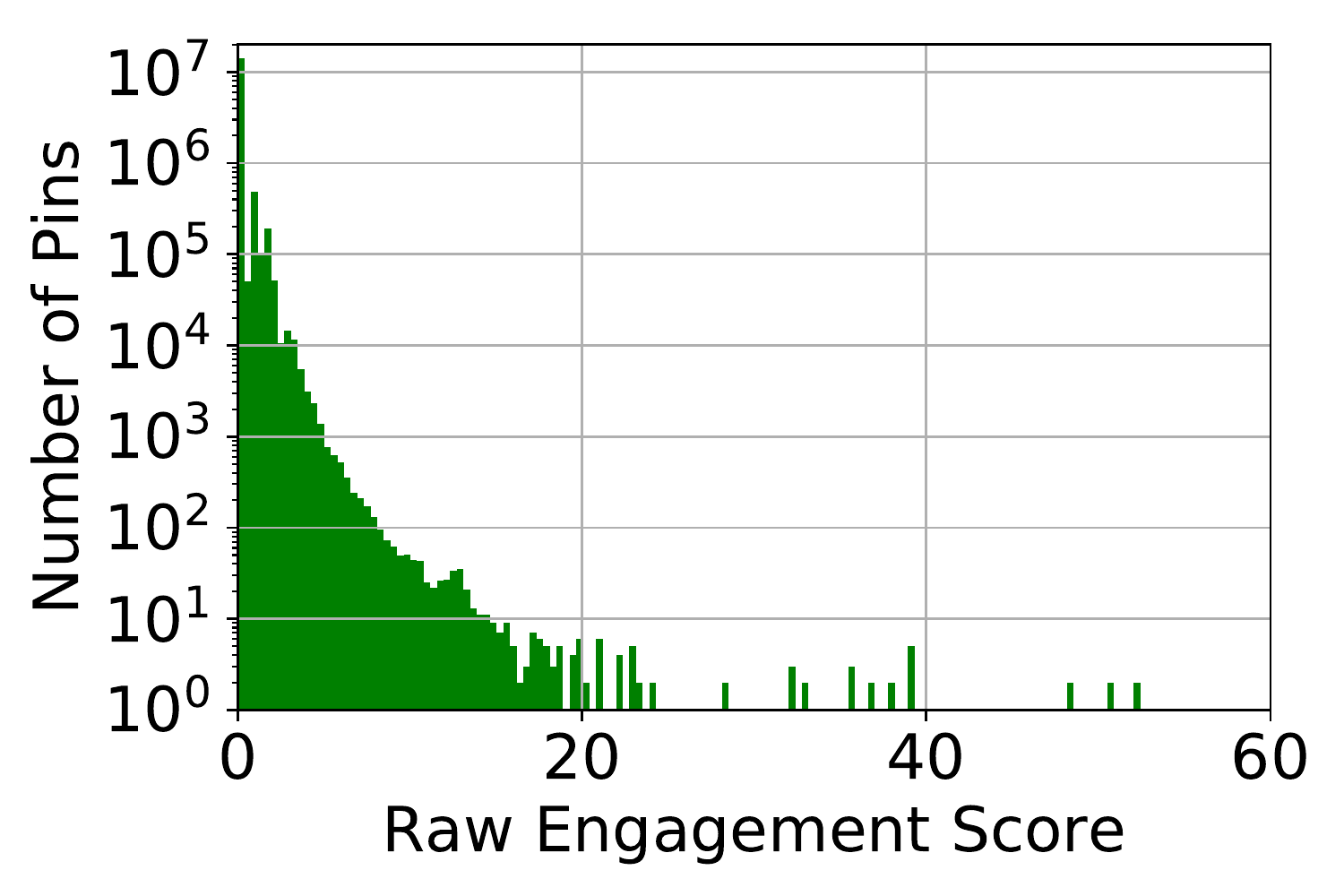}}
\subfigure[Distribution of relevance scores]{\includegraphics[width=0.495\columnwidth]{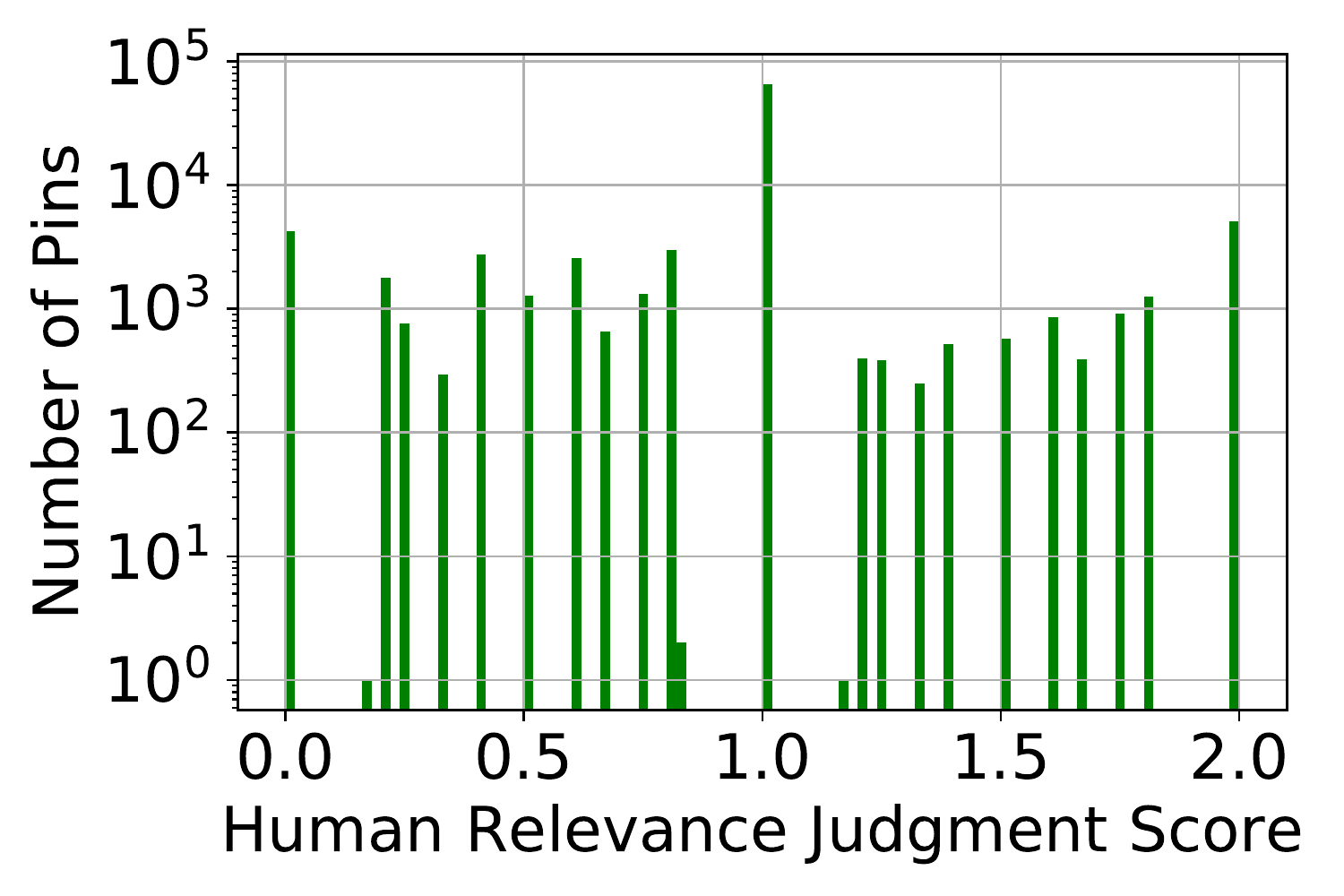}}
\caption{Distribution of quality label $l_p$ across different data sources}\label{fig:score_dist}
\vspace{-0.3cm}
\end{figure}

Clearly, the range of the raw quality label $l_p$ of the human relevance data differs a lot from that of the engagement data. Figure~\ref{fig:score_dist} reports the distribution of quality labels in a set of sampled engagement data and that of human judgment scores in human relevance data after downsampling the negative tuples. Even if we normalize both of them into the same range such as $[0, 1]$, it is still not an apple-to-apple comparison. Therefore, we simply consider each training data source independently and feed each of which into the ranking function to train a specific model and then perform model ensemble in Section~\ref{subsec:ensemble}. This ad-hoc solution performs best in both of our offline and online A/B test evaluation.

\section{Feature Representation for Ranking}\label{subsec:feature}
There are several major groups of features in traditional search engines, which, when taken together, comprise thousands of features~\cite{chapelle2011yahoo}~\cite{geng2007feature}. Here we restrict our discussion to how we enhance traditional ranking features to address unique challenges in Pinterest image search.

\subsection{Beyond Text Relevance Feature}
As discussed earlier, the text description of each Pin usually is very short and noisy. To address this issue, we build an intensive pipeline that generate high-quality text annotations of each pin in the format of unigrams, bigrams and trigrams. The text annotations of one pin are extracted from different sources such as title, description, texts from the crawled linked web pages, texts extracted from the visual image and automatically classified annotation label. These aggregated annotations are thus utilized to compute the text matching score using BM25~\cite{Robertson:2009:PRF:1704809.1704810} and/or proximity BM25~\cite{song2008viewing}.

Even with the high quality image annotation, the text signal is still much weaker and noisier than that in the traditional web page search. Therefore, in addition to word-level relevance measurement, a set of intent-based and embedding-based similarity measurement features are developed to enhance the traditional text-based relevancy.

\begin{description}
\item[Categoryboost] This type of feature tries to go beyond similarity at the word level and compute similarity at the category level. Note that in Pinterest, we have a very precise human curated category taxonomy, which contains 32 L1 categories and 500 L2 categories. Both queries and pins were annotated with categories and their confidences through our multi-label categorizer. 

\item[Topicboost] Similar to categoryboost, this type of feature tries to go beyond similarity at the word level and compute similarity at the topic level. However, in contrast to the category, each topic here denotes a distribution of words discovered by the statistical topic modeling such as Latent Dirichlet allocation topic modeling~\cite{blei2003latent}. 

\item[Embedding Features] The group of embedding features evaluates the similarity between users' query request and the pins based on their distances on the learned distributed latent representation space. Here both word embedding~\cite{mao2016training} and visual embedding~\cite{jing2015visual}~\cite{liu2017related} are trained and inferred via different deep neural network architectures on our own Pinterest Image Corpora.
\end{description}
Our enhanced text relevance features play very important roles in our ranking model. For instance, the categoryboost feature was the 15th important feature in organic search ranking model and was ranked as 1st in search ads relevance ranking model.

\subsection{User Intent Features}
We derive a set of user-intent based features from explicit feedbacks that received from user engagement. 
\begin{description}
\item[Navboost] Navboost is our signal into how well a pin performs in general and in context of a specific query and user segment. It is based on the projected close up, click, long-click and repin propensity estimated from previous user engagement. In addition to segmented signal in terms of types of actions, we also derive a family of Navboost signals segmented by country, gender, aggregation time (e.g., 7 days, 90 days, two years etc).

\item[Tokenboost] Similarly, in order to increase the coverage, another feature Tokenboost is proposed to evaluate how well a pin performs in general and in context of a specific token. 
\item[Gender Features] Pinterest currently has a majority female user base. To ensure we provide equal quality content to male users, we developed a family of gender features to determine, generally, whether a pin is gender neutral or would resonate with men. We then can rank more gender neutral or male-specific Pins whenever a male user searches. For example, if a man searches shoes, we want to ensure he finds shoes for him, not women's shoes.

\item[Personalized Features] As our mission is to help you discover and do what you love, we always put users first and provide as much personalization in results as possible. In order to do this, we rely on not only the demographical information of users, but also various intent-based features such as categories, topics, and embedding of users.
\end{description}
User intent features are one of the most important features for core ranking and they help our learning algorithm learn which type of pins are ``really'' relevant and interesting to users. For instance, the Navboost feature is able to tell the ranking function that a pin about ``travel guides to China '' is much more attractive than a pin about ``China Map'' (which is ranked 1st in Google Image Search) or ``China National Flag'' when a user is searching a query ``China'' in Pinterest.
\subsection{Query Intent Features}
Similar to traditional web search, we also utilize common query-dependent features such as length, frequency, click-through rate of the query. In addition to those common features, we further develop a set of Pinterest-specific features such as whether the query is male-oriented, the ratio between click-through and repin, the category and other intents of queries, and etc. 
\subsection{Socialness, Visual and other Features}
In addition to the above features, there exists more unique features in Pinterest ecosystem. Since each ranking item is an image, dozens of visual related features are developed ranging from simple image score based on image size, aspect ratio to image hashing features.  

Meanwhile, in addition to image search, Pinterest also provide other social products such as image sharing, friends/pin/board following, and cascading image feed recommendation. These products also provide very valuable ranking features such as the socialness, popularity, freshness of a pin or a user etc. 

\section{Cascading Ranking Models}\label{sec:ranking}
Pinterest Search handles billions of queries every month and helps hundreds of millions of monthly active users discover useful ideas through high quality Pins. Due to the huge volume of user queries and pins, it is critical to provide a ranking solution that is both effective and efficient. In this section, we provide a deep-dive walk through of our cascading core ranking module.
\subsection{Overview of the Cascading Ranking}
\begin{figure}[!t]
\centering
\includegraphics[width=\columnwidth]{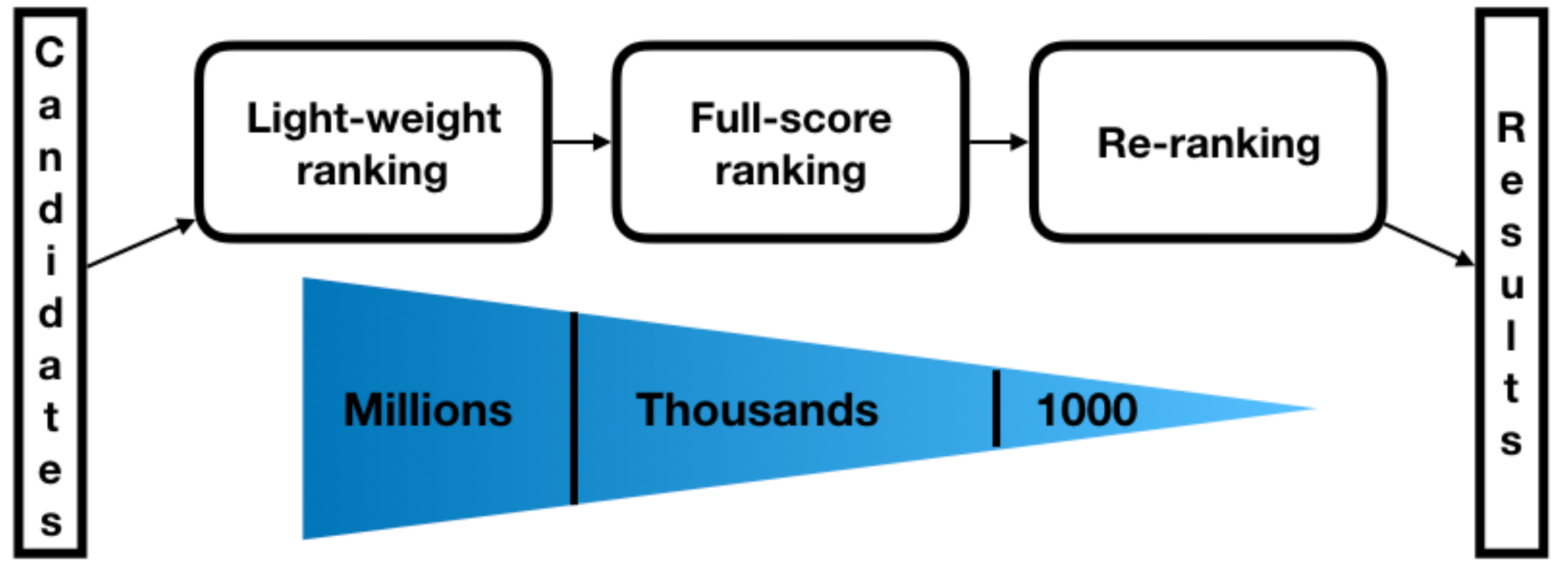}
\caption{An illustrative view of cascading ranking}\label{fig:ranking}
\vspace{-0.3cm}
\end{figure}
As illustrated in Figure~\ref{fig:ranking}, we develop a three-stage cascading ranking module: light-weight stage, full-ranking stage, and re-ranking stage. Note that multi-stage ranking was proposed as early as in \texttt{NestedRanker}~\cite{matveeva2006high} to obtain high accuracy in retrieval. However, only recently motivated by the advances of cascading learning in traditional classification and detection~\cite{raykar2010designing}, cascading ranking~\cite{Liu:2017:CRO:3097983.3098011} has been re-introduced to improve both the accuracy and the efficiency of ranking systems. Coincidently, the Pinterest Image Search System applied a similar cascading ranking design to that of the Alibaba commerce search engine~\cite{Liu:2017:CRO:3097983.3098011}. In the light-weight stage, an efficient model (e.g., linear model) is applied over a set of important but cheaply computed features to filter out negative pins before passing to the full-ranking stage. As shown in Figure~\ref{fig:ranking}, light-weight stage ranking successfully filters out millions of pins and restricts the candidate size for full-ranking to thousands scale. In the full-ranking stage, we select a set of more precise and expensive features, together with a complex model, and further following the model ensemble, to provide a high quality ranking. Finally, in the re-ranking stage, several post-processing steps are applied before returning results to the user to improve freshness, diversity, locale- and language-awareness of results. 

To ease the presentation, we use $q$, $u$, $p$ to denote query, user and pin respectively. $x$ denotes the feature representation for a tuple with query $q$, user $u$ and pin $p$ (see Section~\ref{subsec:feature} for more details). $l(p \mid q, u)$ is the observed quality score of pin $p$ given query $q$ and user $u$, usually is obtained from either the search log or human judgment (see Section~\ref{sec:data}). $y$ is the ground truth quality label of pin $p$ given query $q$ and user $u$, which is constructed from the observed quality score $l(p \mid q, u)$.  Similar to $l(p \mid q, u)$, we use $s(p \mid q, u)$ to denote the scoring function that estimates the quality score of pin $p$ given query $q$ and user $u$. To shorten the notation, we also simply use $l_p$ to denote $l(p \mid q, u)$ and $s_p$ to denote $s(p \mid q, u)$ when the given query $q$ and user $u$ can be omitted without ambiguity. $\mathcal{L}$ denotes the loss function and $\mathcal{S}$ denotes the scoring function.
\subsection{Ranking Models}\label{subsec:model}
\begin{table}[!t]
\caption{A list of models experimented in different stages of the cascading core ranking.}\label{tab:model-comp}
\begin{tabular}{|c|c|c|c|}
\hline
\hline
\textbf{Stage}&Feature&\textbf{Model}&\textbf{Is Pairwise?}\\
\hline
\multirow{2}{*}{Light-weight}&\multirow{2}{*}{8 features}&\texttt{Rule-based}&--\\
\cline{3-4}
&&\texttt{RankSVM}~\cite{joachims2002optimizing}&Pairwise\\
\hline
\hline
\multirow{6}{*}{Full}&\multirow{6}{*}{All features}&\texttt{GBDT}~\cite{li2008mcrank}~\cite{yin2016ranking}&Pointwise\\
\cline{3-4}
&&\texttt{DNN}&Pointwise\\
\cline{3-4}
 &&\texttt{CNN}&Pointwise\\
\cline{3-4}
&&\texttt{RankNet}~\cite{BurgesRankNet2005,burges2010ranknet}&Pairwise\\
\cline{3-4}
&&\texttt{RankSVM}~\cite{joachims2002optimizing}&Pairwise\\
\cline{3-4}
&&\texttt{GBRT}~\cite{zheng2007regression}~\cite{zheng2008general}&Pairwise\\
\hline
\hline
\multirow{4}{*}{Re-ranking}&\multirow{4}{*}{6 features}&\texttt{Rule-based}&--\\
\cline{3-4}
&&\texttt{GBDT}~\cite{li2008mcrank}~\cite{yin2016ranking}&Pointwise\\
\cline{3-4}
&&\texttt{GBRT}~\cite{zheng2007regression}~\cite{zheng2008general}&Pairwise\\
\cline{3-4}
&&\texttt{RankSVM}~\cite{joachims2002optimizing}&Pairwise\\
\hline
\end{tabular}
\vspace{-0.25cm}
\end{table}
As shown in Table~\ref{tab:model-comp}, we experimented a list of representative state-of-the-art models  with our own variation of loss functions and architectures in different stages of the cascading core ranking. In the following, we briefly introduce how we adopt each model into our ranking framework. We omitted the details of the Rule-based model since it is applied very intuitively.

\noindent\textbf{Gradient Boost Decision Tree (GBDT)} Given a continuous and differentiable loss function $\mathcal{L}$, Gradient Boost Machine~\cite{friedman2001greedy} learns an additive classifier $H^T = \sum_{t = 1} ^T \eta_t h^t(x)$ that minimizes $\mathcal{L}(H^T)$, where $\eta$ is the learning rate. In the pointwise setting of \texttt{GBDT}, each $h^t$ is a limited depth regression tree (also referred to as a weak learner) added to the current classifier at iteration $t$. The
weak learner $h^t$ is selected to minimize the loss function $\mathcal{L}(H^{t - 1} + \eta_t h^t)$. We use mean square loss as the training loss for the given training instances:
\begin{equation}
\mathcal{L}(h^t) = \frac{1}{n}\sum_{x, y} (y - h^t(x))^2 
\end{equation}
where $n$ is number of training instances and the ground truth label $y$ is equal to the observed continuous quality label $l(p \mid q, u)$.

\begin{figure}[!t]
\centering
\subfigure[Simple neural network]{\includegraphics[width=0.55\columnwidth]{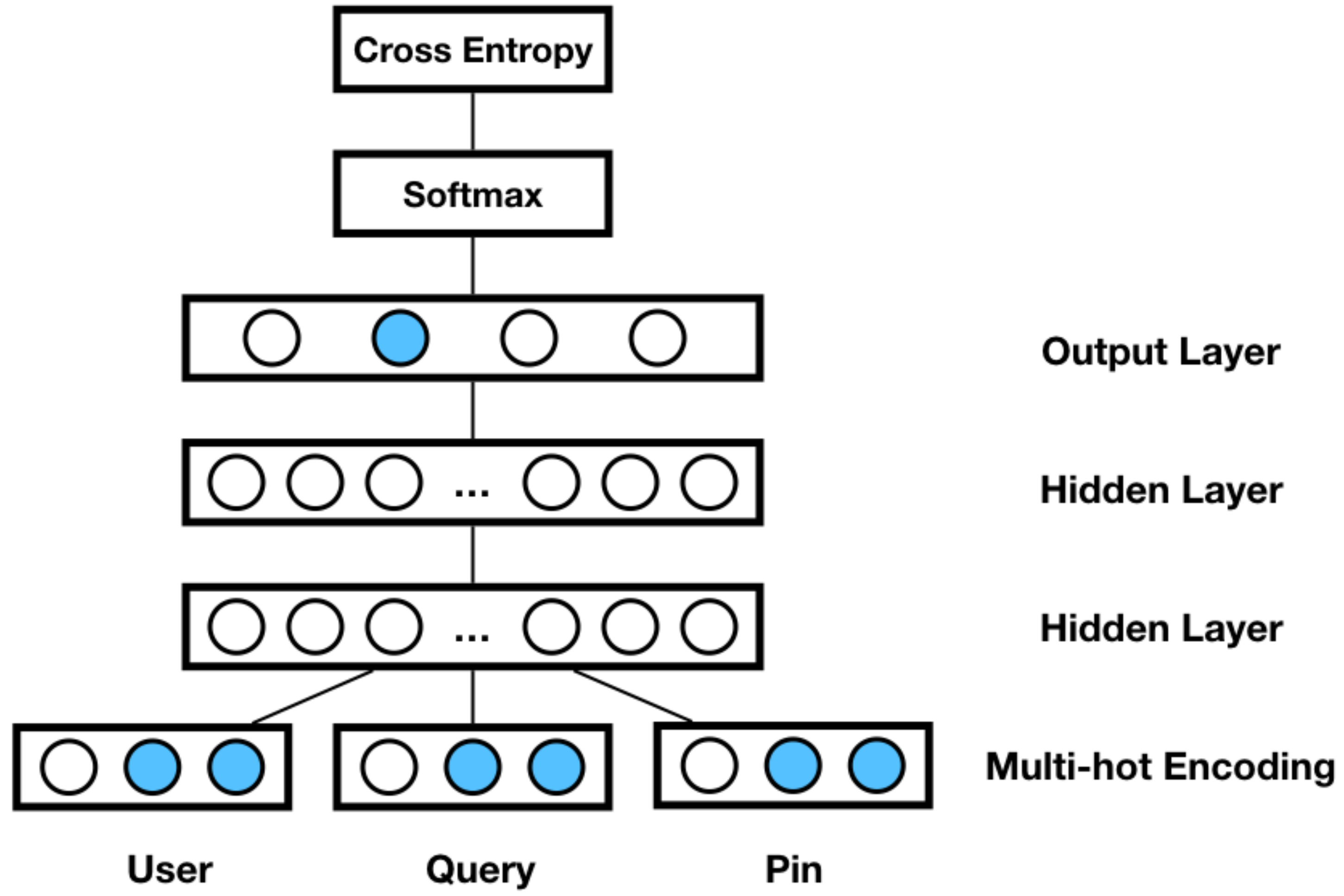}}
\subfigure[convolutional neural network]{\includegraphics[width=0.55\columnwidth]{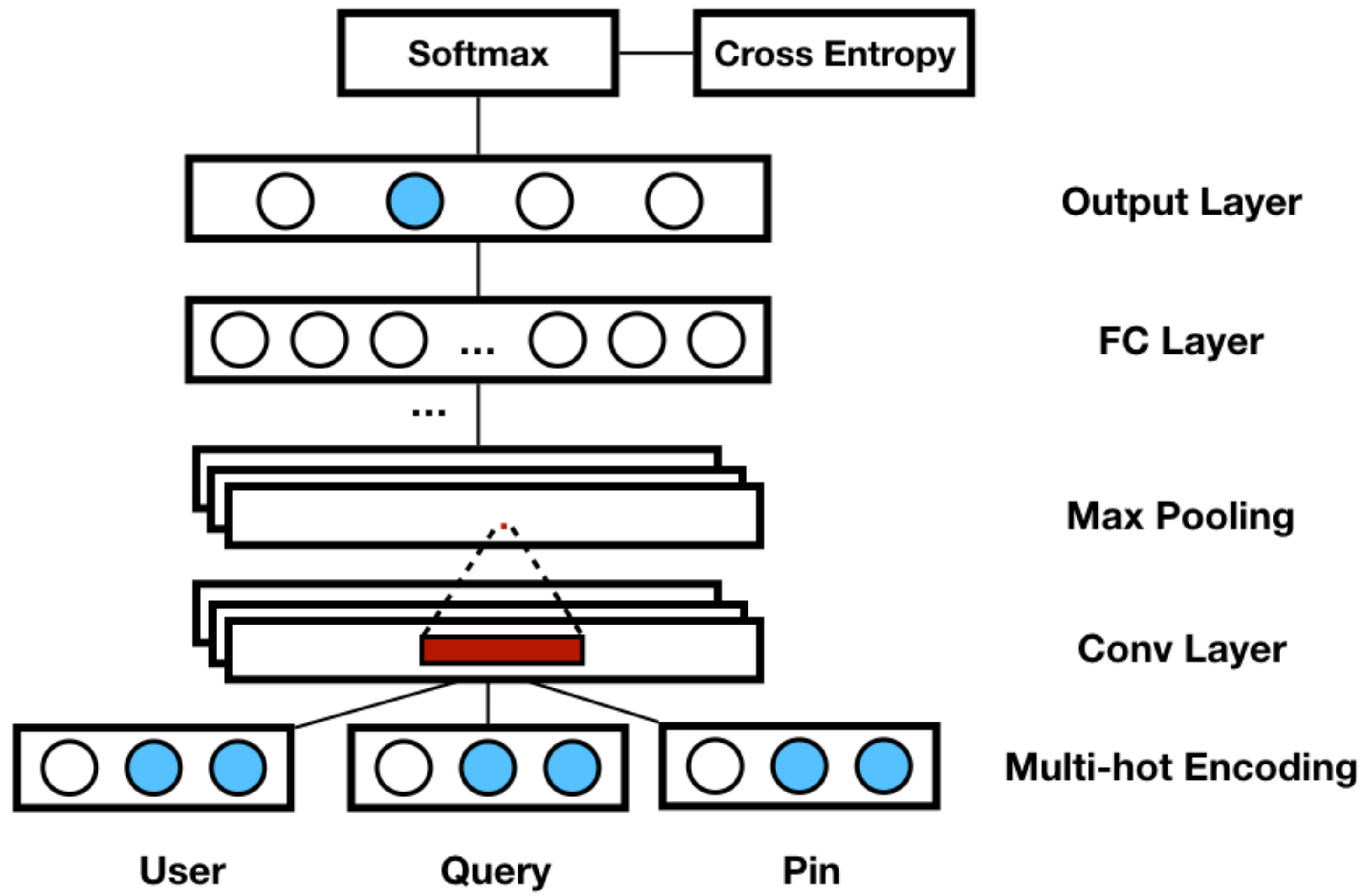}}
\caption{Different ranking architectures}\label{fig:NN}
\vspace{-0.25cm}
\end{figure}

\noindent\textbf{Deep Neural Network (DNN)} The conceptual architecture of the \texttt{DNN} model is illustrated in Figure~\ref{fig:NN}(a). This architecture models a point-wise ranking model that learns to predict quality score $s(p \mid q, u)$.

Instead of directly learning a scoring function $\mathcal{S}(q, u, p \mid \theta)$ that determines the quality score of pin $p$ for query $q$ and user $u$ given a set of model parameters $\theta$~\cite{dehghani2017neural}, we transform the problem into a multi-class classification problem that classifies each pin into a $4$-scale label [1, 2, 3, 4]. Specifically, during the training phase, we discretize the continuous quality label $l(p \mid q, u)$ into the ordinal label $y \in [1, 2, 3, 4]$ and train a multi-class classifier $\mathcal{S}(k \mid q, u, p, \theta)$ that predicts the probability of pin $p$ in class $k$. 

As shown in Figure~\ref{fig:NN}(a), we use cross entropy loss as the training loss for a single training instance:
\begin{equation}
\mathcal{L} (\mathcal{S}, y) = - \sum_{k =1} ^{K} \mathbf{1}\{y = k\} \log \mathcal{S}(k \mid q, u, p, \theta)
\end{equation}
where $K$ is number of class labels ($K$ = 4 in this setting).

In the inference phase, we treat the trained model as a point-wise scoring function to score each pin $p$ for query $q$ and user $u$ using the following conversion function:
\begin{equation}\label{eq:score}
s(p \mid q, u)= \sum_{k} k * \mathcal{S}(k \mid q, u, p, \theta)
\end{equation}

\noindent\textbf{Convolutional Neural Network (CNN)} In this model, similar to the previous \texttt{DNN} model, the goal is to learn a multi-class classifier $\mathcal{S}(k \mid q, u, p, \theta)$ and then convert the predicted probability of $\mathcal{S}(k \mid q, u, p, \theta)$ into a scoring function $s(p \mid q, u)$ using Eq.~\ref{eq:score}. As it is depicted in Figure~\ref{fig:NN}(b), the architecture contains the $1^{\texttt{st}}$ layer of convolutional layer, following the max pooling layer, with the ReLU activator, the $2^{\texttt{nd}}$ layer of convolutional layer, again following the max pooling layer and the ReLU activator, a fully connected layer and the output layer. 

Despite the differences in the architecture, the \texttt{CNN} model uses the same problem formulation, cross entropy loss function, and score conversion function (Eq.~\ref{eq:score}) as the \texttt{DNN}.

\noindent\textbf{RankNet} Burges et. al.~\cite{BurgesRankNet2005} proposed to learn ranking using a probabilistic cost function based on pairs of examples. Intuitively, the pairwise model tries to learn the correct ordering of pairs of documents in the ranked lists of individual queries. In our setting, one model learns a ranking function $\mathcal{S}(q, u, p_i, p_j, \theta)$ which predicts the probability of pin $p_i$ to be ranked higher than $p_j$ given query $q$ and user $u$. 

Therefore, in the training phase, one important tasks is to extract the preference pair set $\mathcal{P}$ given query $q$ and user $u$. In \texttt{RankNet}, the preference pair set was extracted from the pairs of consecutive training samples in the ranked lists of individual queries. When applying \texttt{RankNet} to our Pinterest search ranking, the preference pair set is constructed based on the raw quality label $l(p\mid q, u)$. For instance, $p_i$ is preferred over $p_j$ if $l(p_i\mid q, u) > l(p_j\mid q, u)$.  Note that the preference pair set construction is applied to all the following pairwise models.

Given a preference pair ($p_i$, $p_j$), Burges et. al.~\cite{BurgesRankNet2005} used the cross entropy as the loss function in \texttt{RankNet}:
\begin{equation}\label{eq:ranknetloss}
\mathcal{L}(\mathcal{S}, y) =  - y_{ij}\log \mathcal{S} (q, u, p_i, p_j, \theta) - (1 - y_{ij}) \log (1- \mathcal{S} (q, u, p_i, p_j, \theta))
\end{equation}
where $y_{ij}$ is the ground truth probability of pin $p_i$ ranked higher than $p_j$.

The model was named as \texttt{RankNet} since Burges et. al.~\cite{BurgesRankNet2005} used a two-layer Neural Network to optimize the loss function in Eq.~\ref{eq:ranknetloss}. The very recent rank model proposed by Dehghani et. al.~\cite{dehghani2017neural} can be considered as a variant of \texttt{RankNet}, which used Hinge loss function and a different way of converting the pairwise ranking probability into a scoring function.

\noindent\textbf{RankSVM} In the pairwise setting of \texttt{RankSVM}, given the preference pair set $\mathcal{P}$, \texttt{RankSVM}~\cite{joachims2002optimizing} aims to optimize the following problem:
\begin{equation}
\argmin_{w}\frac{1}{2}\|w\|^2 + c \sum_{i}\sum_{j, k \in \mathcal{P}_i} \mathcal{L}(w^T x_j - w^T x_k)
\end{equation}
A popular loss function used in practical is the quadratically smoothed hinge loss~\cite{zhang2004solving} such that $\mathcal{L}(\epsilon) = \max(0, 1 - \epsilon)^2$.

\noindent\textbf{Gradient Boost Ranking Tree (GBRT)} Intuitively, one can weigh the \texttt{GBRT} as a combination of \texttt{RankSVM} and \texttt{GBDT}. In the pairwise setting of \texttt{GBRT}, similar to \texttt{RankSVM}, at each iteration the model aims to learn a ranking function $\mathcal{S}(q, u, p_i, p_j, \theta)$ that predicts the probability of pin $p_i$ to be ranked higher than $p_j$ given query $q$ and user $u$. In addition, similar to the setting of \texttt{GBDT}, here the ranking function is a limited depth regression tree $h^t$. Again, the decision tree $h^t$ is selected to minimize the loss $\mathcal{L}(H^{t - 1} + \eta_t h^t)$, where the loss function is defined as:
\begin{equation}
\mathcal{L}(h^t) = \sum_{i}\sum_{j, k \in \mathcal{P}_i} \max (0, h^t(x_k) - h^t(x_j) + \epsilon)^2 
\end{equation}

\subsection{Model Ensemble across Different Data Sources}\label{subsec:ensemble}
In this section, we discuss how we perform calibration over multiple models that are trained from different data sources (e.g., engagement training data versus human relevance data).

Various ensemble techniques~\cite{dietterich2000ensemble} are proposed to decrease variance and bias and improve predictive accuracy such as stacking, cascading, bagging and boosting (\texttt{GBDT} in Section~\ref{subsec:model} is a popular boosting method). Note that the goal here is not only to improve the quality of ranking using multiple data sources, but also to maintain the low latency of the entire core ranking system. Therefore, we here consider a specific type of ensemble approach \textbf{stacking} with relatively low computational costs.

Stacking first trains several models from different data sources and the final prediction is the linear combination of these models. It introduces a meta-level and uses another model or approach to estimate the weight of each model, i.e., to determine which model performs well given these input data. 

Note that stacking can be performed both within the training of each individual model or after the training of each individual model. When stacking is applied after training each individual model, then the final scoring function is defined as
\begin{equation}
s(p \mid q, u) = \gamma s_e (p\mid q, u) + (1 - \gamma) s_r(p\mid q, u) 
\end{equation}
where $s_e$/$s_r$ is the predicted score of the model from engagement/human relevance judgment data and $\gamma$ is the combination coefficient.

Stacking can also be performed within model training. For instance, Zheng et. al.~\cite{zheng2008general} linearly combined the tree model that fits the engagement data and another tree model that fits the human judgment data using the following loss function:
\begin{equation}\label{eq:stackloss}
\mathcal{L}(h^t) = \gamma \sum_{i}\sum_{j, k \in \mathcal{P}_i} \max (0, h^t(x_k) - h^t(x_j) + \epsilon)^2  + (1 - \gamma) \sum_{i} (y_i - h^t(x_i))^2
\end{equation}
where $y_i$ is the relevance label for pin $i$ and $\gamma$ controls the contribution of each data source.

Here we chose to perform stacking at different stages based on the complexity of each individual model: stacking is performed in the model training phase if each individual model is relatively easy to compute, and is performed after training each individual model vise versa (e.g., each individual model is a neural network model). 

Note that differs from Eq.~\ref{eq:stackloss}, we always use the same loss function for different data sources. For instance, assume that we aim to train \texttt{GBRT} tree models from both engagement training data and human relevance data, we simply optimize the combined pairwise loss function:
\begin{equation}
\begin{aligned}
\mathcal{L}(h^t) = &\gamma \sum_{i}\sum_{j, k \in \mathcal{P}_i} \max (0, h^t(x_k) - h^t(x_j) + \epsilon)^2 \\
&+ (1 - \gamma) \sum_{n}\sum_{j, k \in \mathcal{P}_n} \max (0, h^t(x_k) - h^t(x_j) + \epsilon)^2
\end{aligned}
\end{equation}
where each $\mathcal{P}_i$/$\mathcal{P}_n$ denotes a preference set extracted from engagement /human judgment data respectively, and $\gamma$ again controls the contribution of each data source. The advantage of this loss function is that $\gamma$ can also be intuitively explained as proportional to number of trees grown from each data source.

\section{Experiment}\label{sec:expt}
\subsection{Offline Experimental Setting}
The first group of experiments was conducted off-line on the training data extracted as described in Section~\ref{sec:data}. For each country and language, we curated 5000 queries and performed human judgment for 400 pins per query. In addition, we built the engagement training data pipeline from randomly extracting recent 7-days 1\% search user session Log. The full data set was randomly divided while 70\% was used for training, 20\% used for testing and 10\% used for validation. In total we have 15 millions of training instances.

\subsubsection{Feature Statistics}
\begin{figure}[htb]
\centering
\subfigure[Text relevance feature]{\includegraphics[width=0.495\columnwidth]{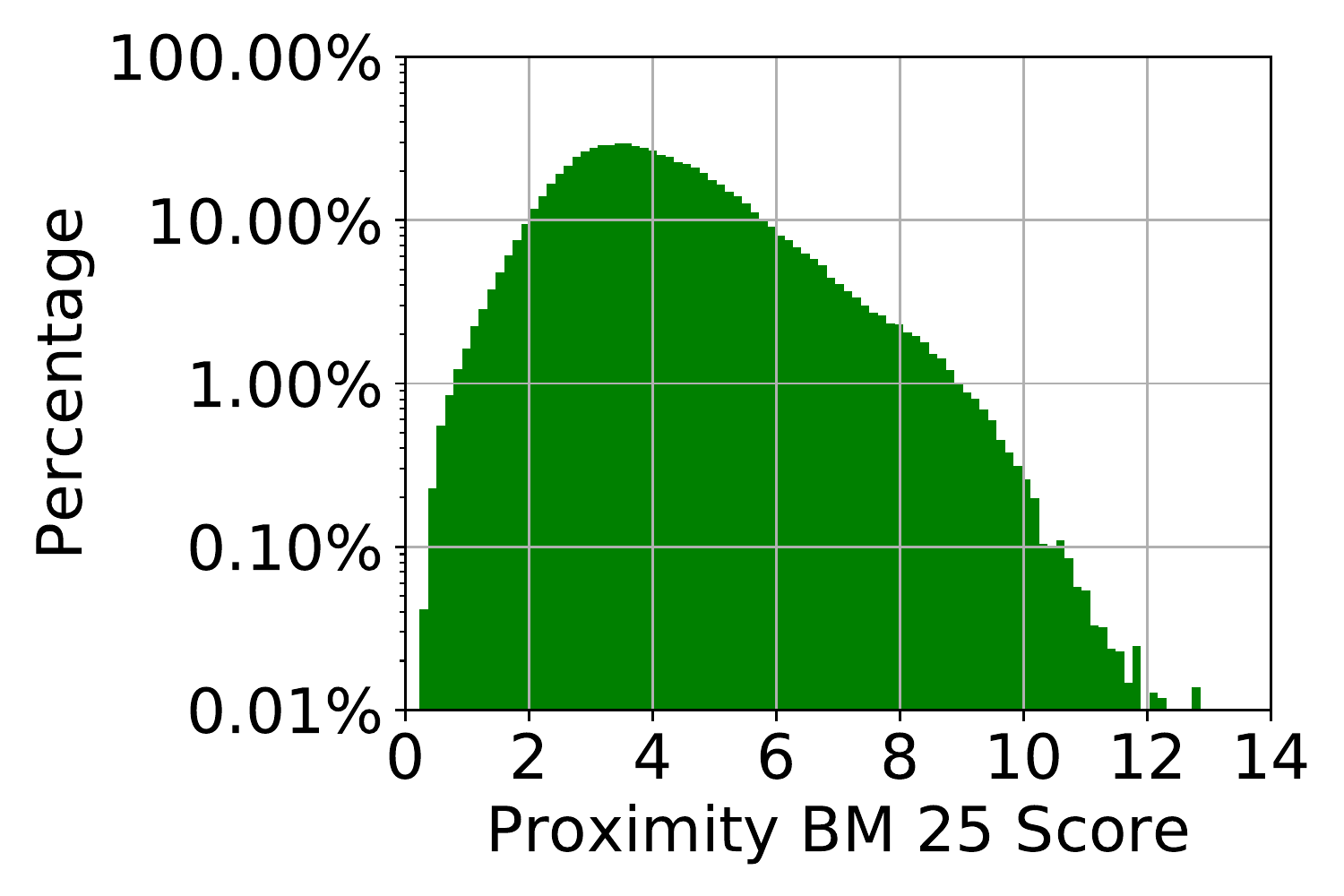}}
\subfigure[Social feature]{\includegraphics[width=0.495\columnwidth]{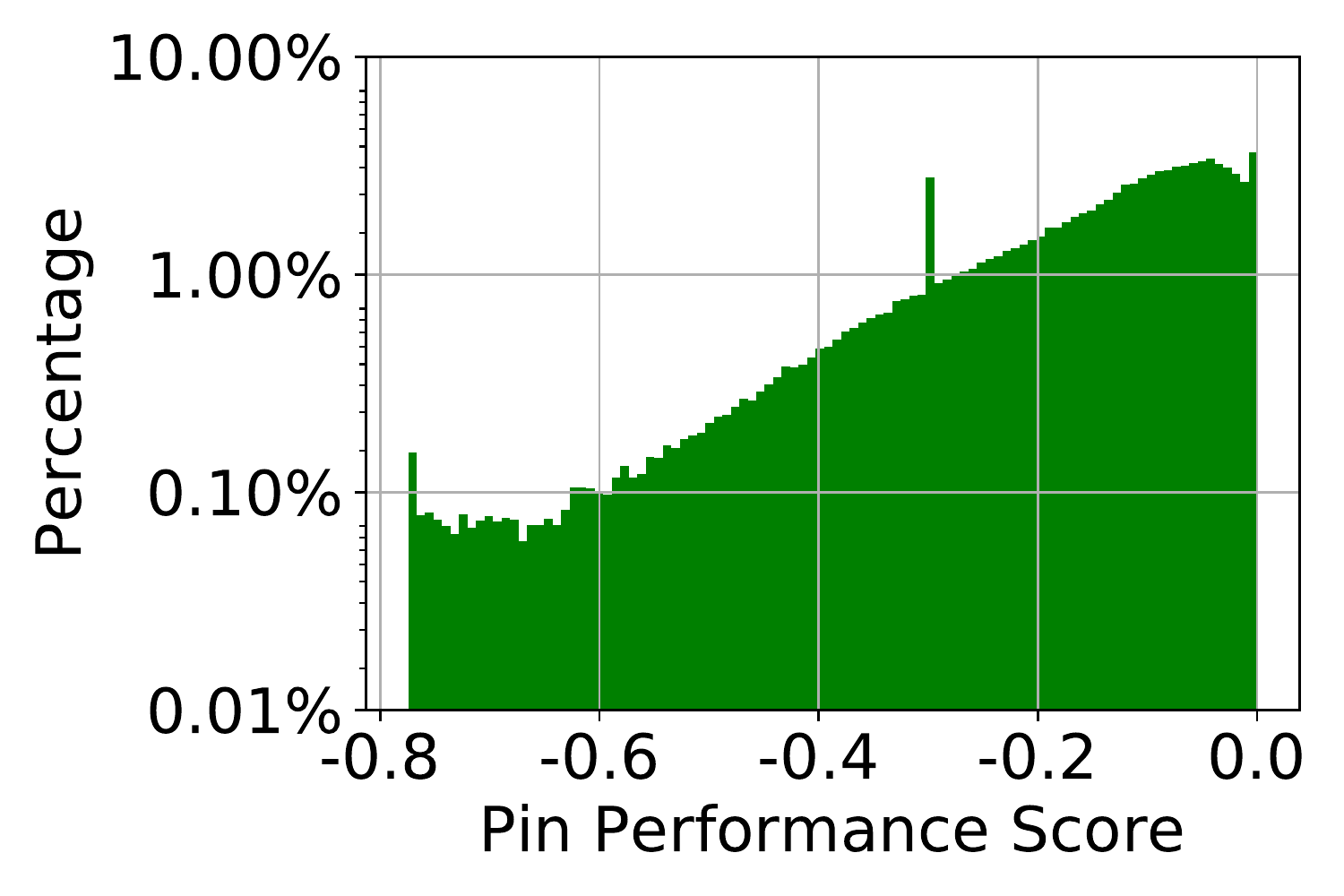}}
\subfigure[Query intent feature]{\includegraphics[width=0.495\columnwidth]{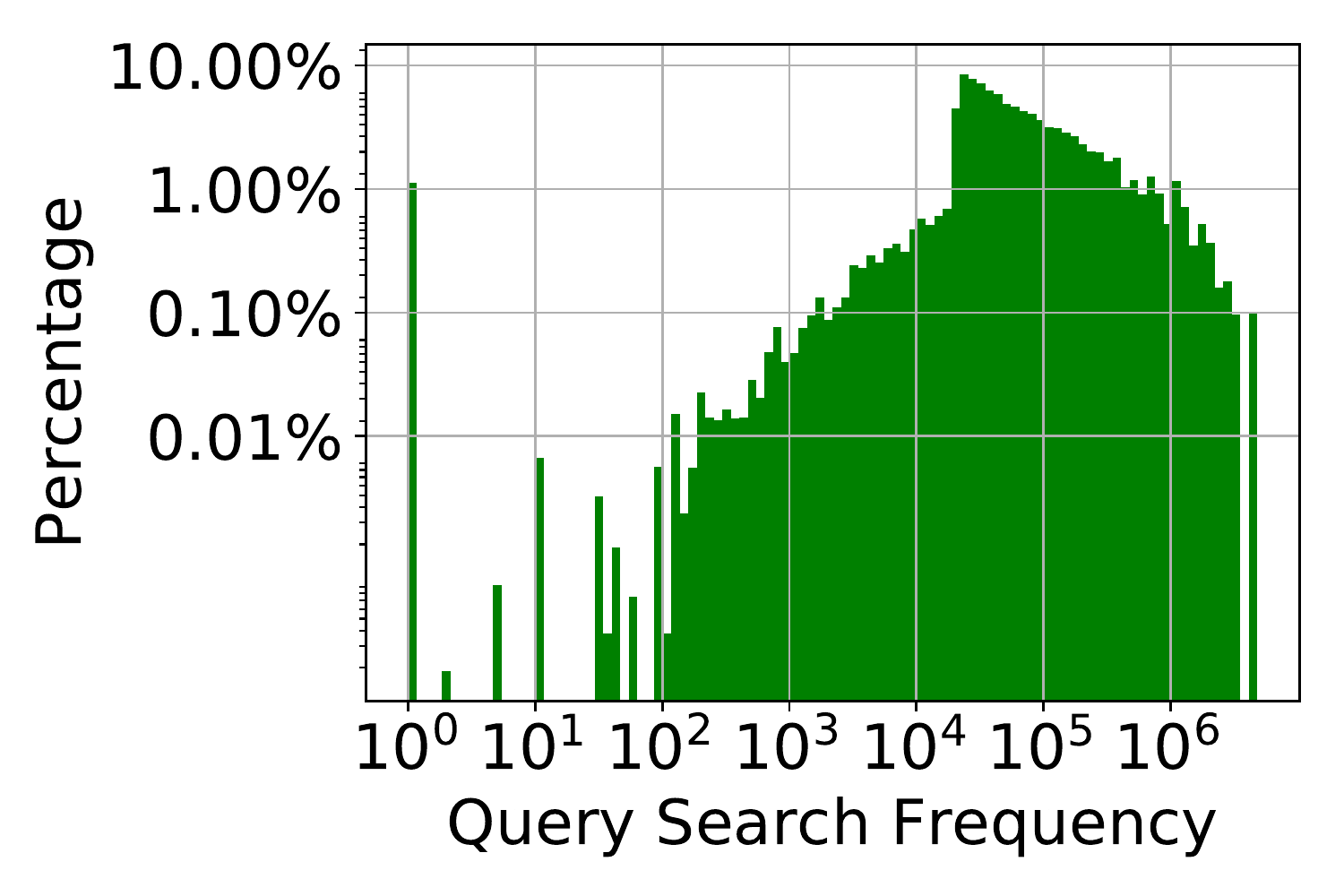}}
\subfigure[User intent feature]{\includegraphics[width=0.495\columnwidth]{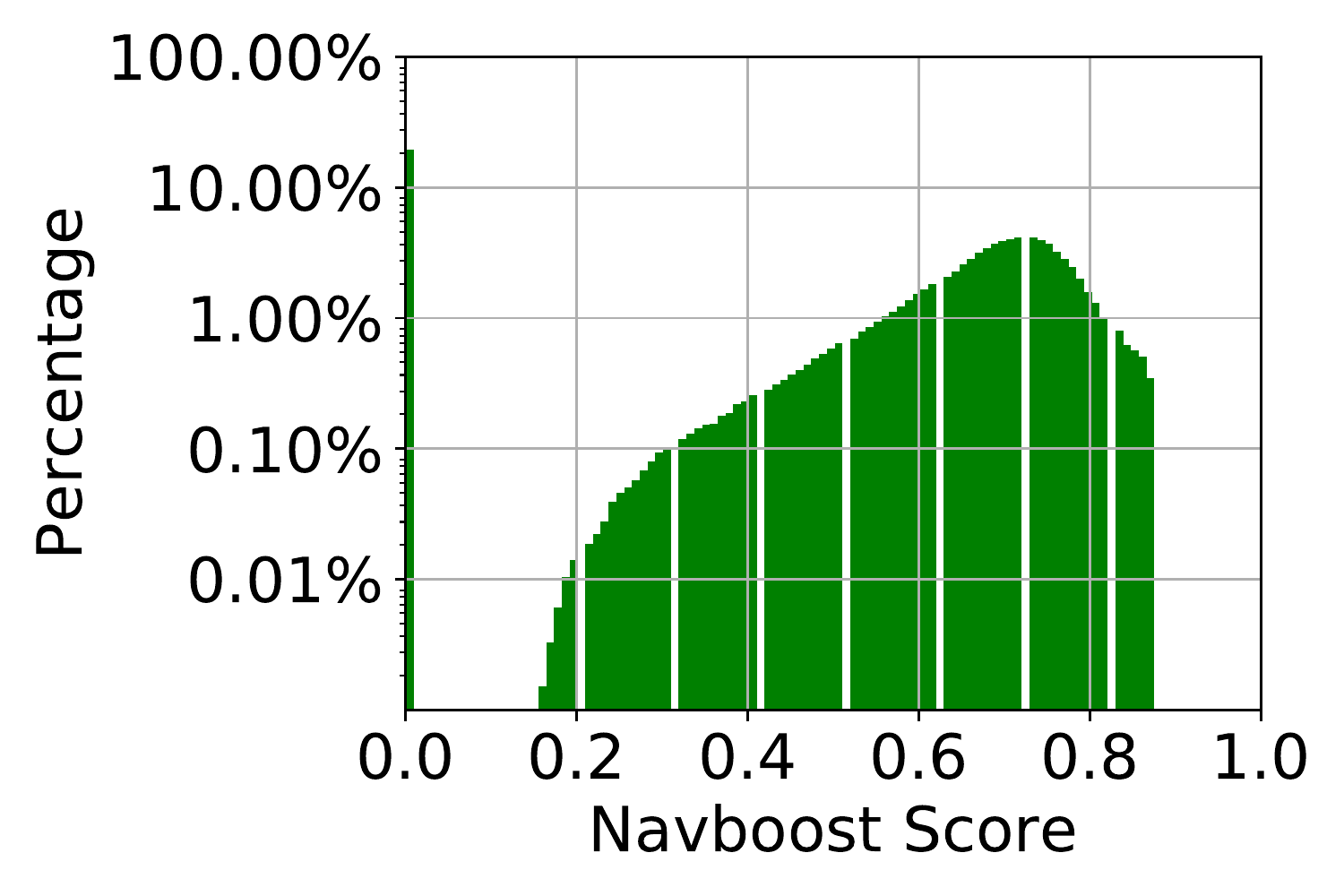}}
\caption{Distribution of selected feature values}\label{fig:feature_dist}
\vspace{-0.4cm}
\end{figure}

We also analyzed the coverage and distribution of each individual feature. Due to the space limitation, we report the statistics of the top important features from each group in Figure~\ref{fig:feature_dist}.

\subsubsection{Offline Measurement Metrics} In offline setting, we use the query-level Normalized Discounted Cumulative Gain (\textbf{NDCG}~\cite{JarvelinNDCG2002}). Given a list of documents and their ground truth labels $l$, the discounted cumulative gain at the position $p$ is defined as:
\begin{equation}
\texttt{DCG}_p = \sum_{r = 1} ^{p} \frac{l_r}{\log (r + 1)}
\end{equation}
The NDCG is thus defined as:
\begin{equation}
\texttt{NDCG}_p = \frac{\texttt{DCG}_p}{\texttt{IDCG}_p}
\end{equation}
where $\texttt{IDCG}_p$ is the ideal discounted cumulative gain.

Since we have two different data sources, we derived two measurement metrics: $\texttt{NDCG}_p^{r}$ for the human relevance data and $\texttt{NDCG}_p^{e}$ for the engagement data. 
\subsection{Online Experimental Setting}
A standard A/B test is conducted online, where users are bucked into different 100 buckets and both the control group and enabled group can use as much as 50 buckets. In this experiment, 5\% users in the control group were using the old in production ranking model, while another 5\% users in the enabled group were using the experimental ranking model. 

The Pinterest image search engine handles in average 2 billion monthly text searches, 600 million monthly visual searches, 70 millions of queries everyday
and the query volume could be doubled during the peak periods such as Valentine's day, Halloween etc. Therefore, roughly 7 millions of queries per day and their search results were evaluated in our online experiments. 

\subsubsection{Online Measurement Metrics}
In online setting, we use a set of both user-level measurement metrics and query-level measurement metrics. For query-level measurement metrics, repin per search ($Q_{\texttt{repin}}$), click per search ($Q_{\texttt{click}}$), close up per search ($Q_{\texttt{close up}}$) and engagement per search ($Q_{\texttt{engaged}}$) were the main metrics we used. This is because repin, click and close up are the main three types among in total 60 types of actions. The volume of close up action (user clicked on any of the pins to see the zoomed in image and the description of pins) is the dominant since this action is the cheapest. To the contrary, the volume of click action is much lower because click is more expensive to act (As shown in Figure~\ref{fig:UI}, the click means that a user clicked the hyperlinks of the pins and went to the external linked web pages after closing up action).

In the user-level, we use the following measurement metrics:
\begin{equation}
\begin{aligned}
U_{\texttt{repin}} = & \frac{\mbox{\# of repined users}}{\mbox {\# of searchers}} & U_{\texttt{close up}} = & \frac{\mbox{\# of close up users}}{\mbox {\# of searchers}}\\
U_{\texttt{click}} = & \frac{\mbox{\# of clicked users}}{\mbox {\# of searchers}} & U_{\texttt{engaged}} = &\frac{\mbox{\# of engaged users}}{\mbox {\# of searchers}}\\
\end{aligned}
\end{equation}

In order to evaluate the effect of re-ranking in terms of boosting local and fresh content, we also use the following measurement metrics:

\begin{equation}
\begin{aligned}
L_{\texttt{imp}} = & \frac{\mbox{\# of local impressed pins}}{\mbox {\# of impressed pins}} & F_{\texttt{imp}} = & \frac{\mbox{\# of fresh impressed pins}}{\mbox {\# of impressed pins}}\\
L_{\texttt{repin}} = & \frac{\mbox{\# of local pins repined}}{\mbox {\# of local pins}} &F_{\texttt{repin}} = & \frac{\mbox{\# of fresh pins repined}}{\mbox {\# of fresh pins}}\\
L_{\texttt{click}} = & \frac{\mbox{\# of local pins clicked}}{\mbox {\# of local pins}} &F_{\texttt{click}} = & \frac{\mbox{\# of fresh pins clicked}}{\mbox {\# of fresh pins}} \\
\end{aligned}
\end{equation}
where local pins denote that the linked country of pins matches that of users, and fresh pins denote the pins with ages no older than 30 days.

\subsection{Performance Results}
\subsubsection{Lightweight Ranking Comparison}
\begin{figure}[!t]
\centering
\subfigure[Offline performance]{\includegraphics[width=0.495\columnwidth]{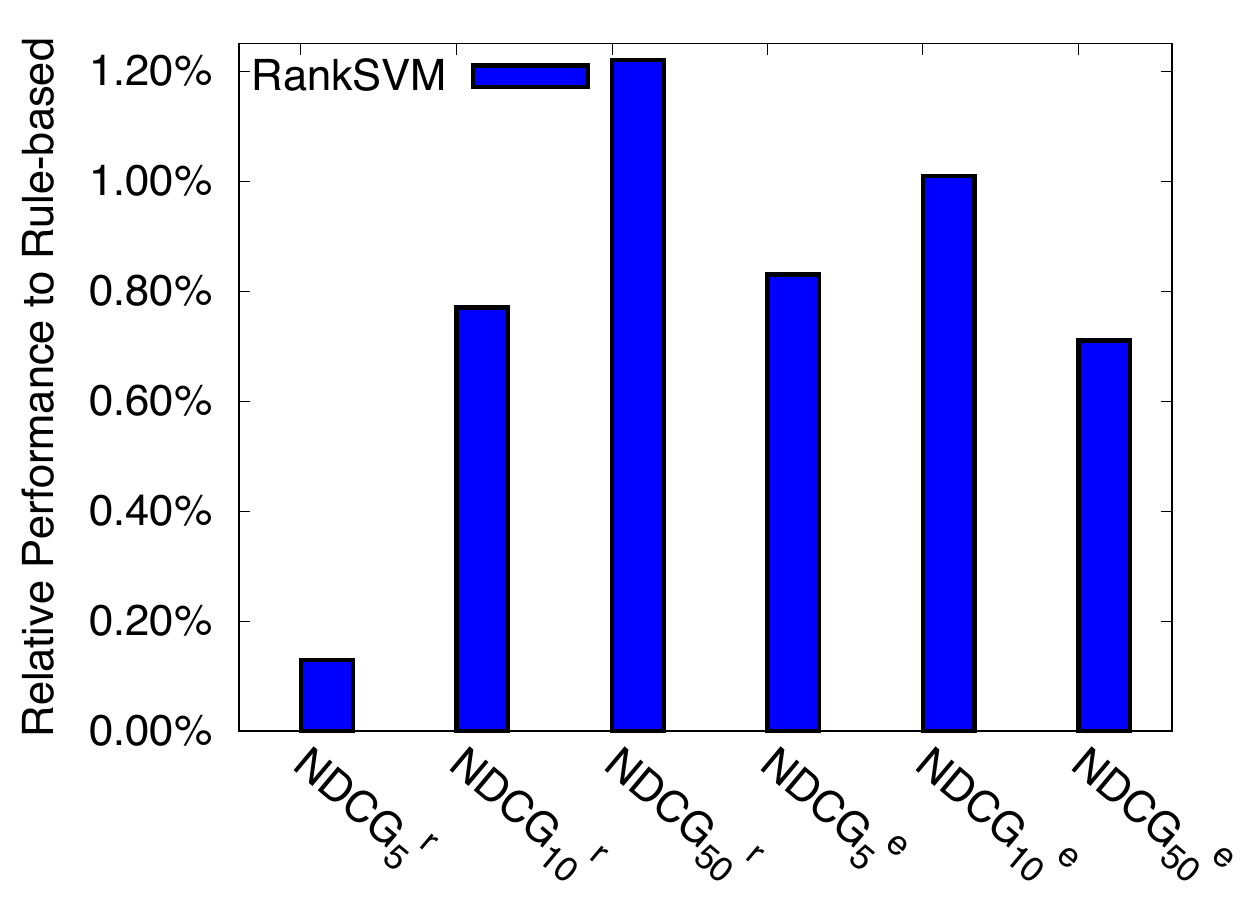}}
\subfigure[Online performance]{\includegraphics[width=0.495\columnwidth]{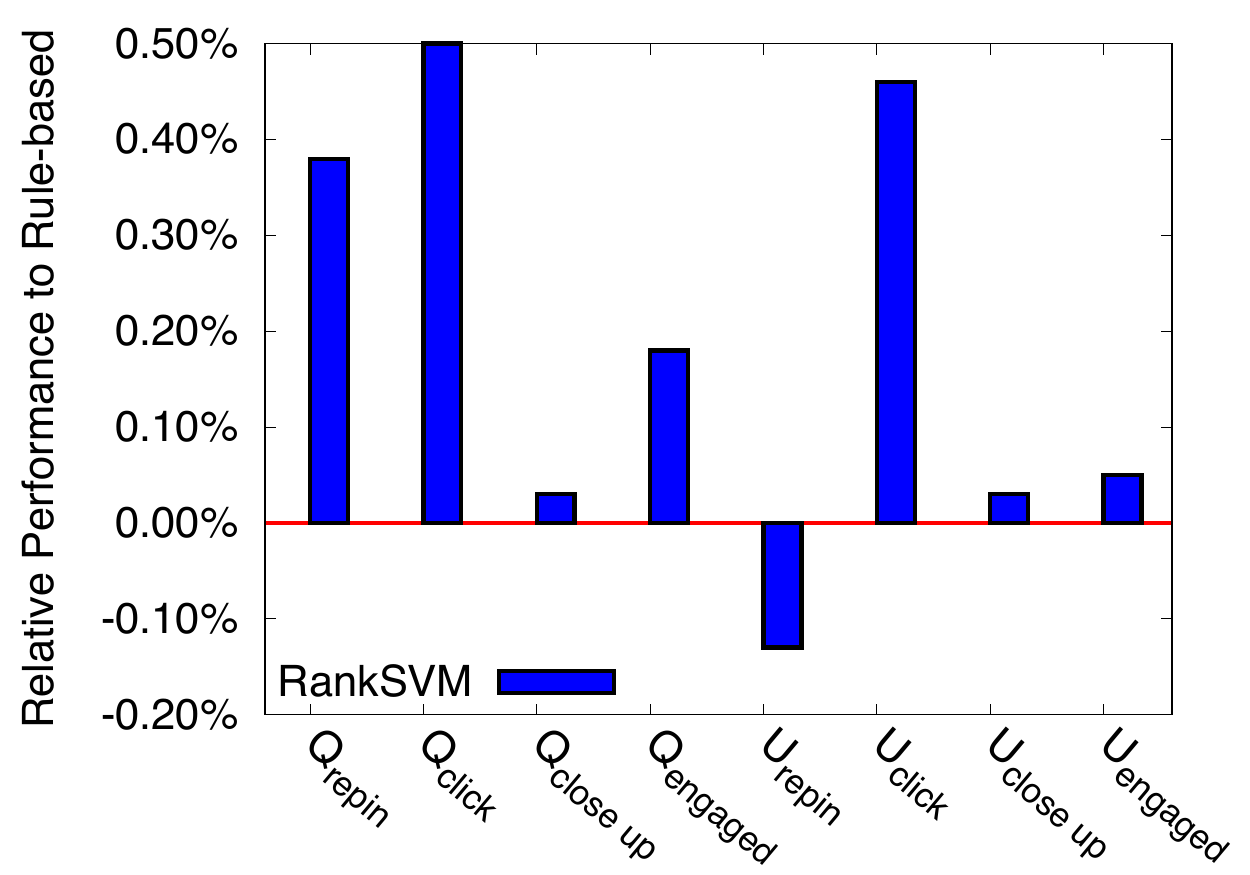}}
\caption{Relative performance of RankSVM model to the baseline rule-based method in lightweight ranking stage.}\label{fig: results-lightweight}
\vspace{-0.25cm}
\end{figure}

\begin{table}[!t]
\caption{Latency Improvement of RankSVM Lightweight Ranking}\label{tab:latency}
\begin{tabular}{|c|c|c|}
\hline
Latency& Rule-based&RankSVM\\
\hline
< 50ms&5\%&8\% \\
\hline
50 - 200 ms&43\%&61\% \\
\hline
> 200 ms &52\%&31\% \\
\hline
\end{tabular}
\vspace{-0.25cm}
\end{table}
The relative performance of \texttt{RankSVM} model to our very earlier rule-based ranking model in lightweight ranking stage is summarized in Figure~\ref{fig: results-lightweight}. In offline test data set, the \texttt{RankSVM} model obtained consistent improvement over the rule-based ranking model. However, when moving to the online A/B test experiment, the improvement is smaller. These phenomena are very consistent across all of the ranking experiments: It is much easier to tune a better model than baseline model in offline than online.

Although the quality improvement is relatively subtle, we greatly reduced the search latency when migrating the rule-based ranking to the \texttt{RankSVM} model. With the \texttt{RankSVM} model in the lightweight stage, we have higher confidences in filtering negative pins before passing the candidates into the full ranking stage. This subsequently improves the latency. As shown in Table~\ref{tab:latency}, the percentage of search latency that is smaller than 50 ms is increased from 5\% to 8\% while the percentage of search latency that is larger than 200 ms is reduced from 52\% to 31\%.

The results reported in Figure~\ref{fig: results-lightweight} and Table~\ref{tab:latency} perfectly illustrated how we achieve the balance between search latency and search quality with the lightweight ranking model. The \texttt{RankSVM} model for the lightweight stage was initially launched and serving all the US traffic starting April 2017.

\subsubsection{Full Ranking Comparison}
\begin{figure}[!t]
\centering
\subfigure[Offline performance]{\includegraphics[width=0.495\columnwidth]{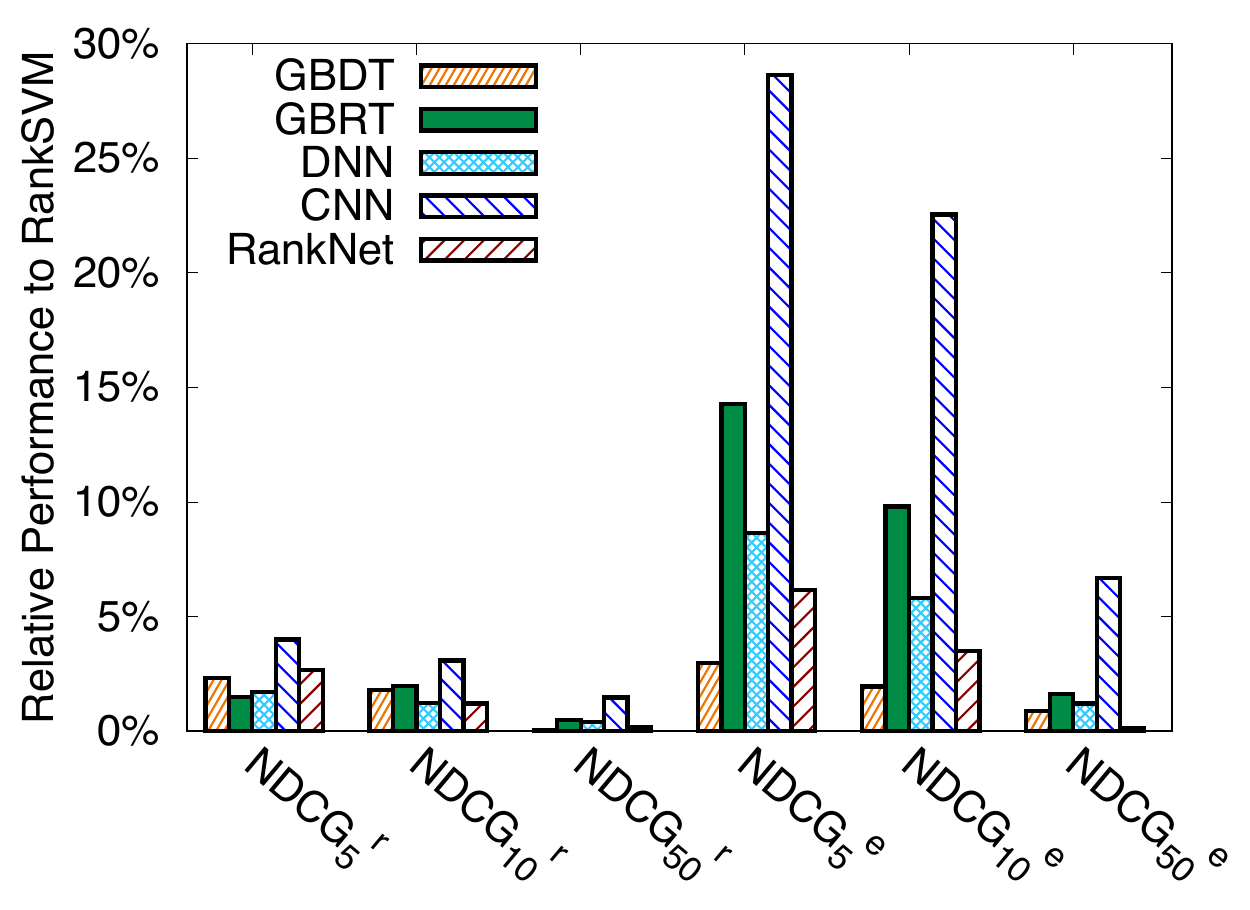}}
\subfigure[Online performance]{\includegraphics[width=0.495\columnwidth]{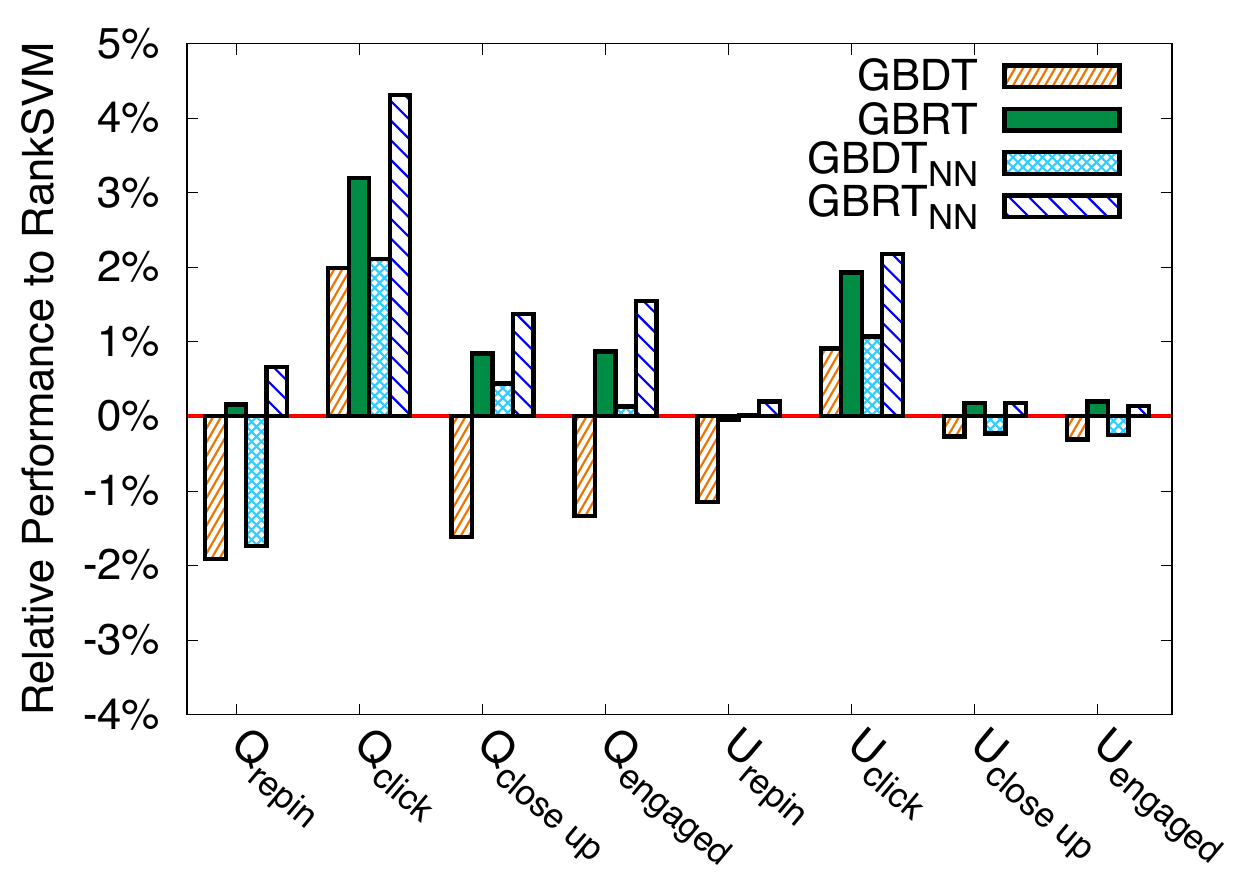}}
\caption{Relative performance of different models to the baseline RankSVM method in full ranking stage. }\label{fig:results-full-ranking}
\vspace{-0.25cm}
\end{figure}

In the full ranking stage, we conduct detailed experiments in off line to compare the performance of different models. As shown in Figure~\ref{fig:results-full-ranking}(a), for the engagement-based quality, overall, \texttt{CNN} $\succeq$ \texttt{GBRT} $\succeq$ \texttt{DNN} $\succeq$ \texttt{RankNet} $\succeq$ \texttt{GBDT}, where $A\succeq B$ denotes A performs significantly better than B. In terms of relevance-based quality, \texttt{CNN} $\succeq$ \{\texttt{GBRT}, \texttt{DNN}, \texttt{RankNet}, \texttt{GBDT}\}.

Although Neural Ranking models perform very well in off line, currently our online model serving platform for neural ranking models incurs additional latency. The latency might be ignorable for recommendation-based products but causes bad experiences for searchers in terms of increased pinner waiting time etc. Therefore, we compute the ranking scores of DNN and CNN models in off line and feed these as two additional features into online tree models, denoted as $GBRT_{\texttt{NN}}$ and $GBDT_{\texttt{NN}}$ respectively. The results of online experiment are presented in Figure~\ref{fig:results-full-ranking}(b). Based on the significant improvement of \texttt{GBRT} over the old linear \texttt{RankSVM} model, we launched the \texttt{GBRT} model in product in October 2017 and will launch the $GBRT_{\texttt{NN}}$ model to serve the entire search traffic soon.
\subsubsection{Re-ranking Comparison}
Note that the main purposes of the re-ranking is to improve the freshness and localness of results. In the early days, our re-ranking applied a very simple hand-tuned rule-based ranking functions. For example, assume that users prefer to see more fresh content, we then simply give any pin with age younger than 30 days a boost or enforce at least a certain percentage of returned results are fresh. 

\begin{figure}[thb]
\centering
\subfigure[Offline performance]{\includegraphics[width=0.495\columnwidth]{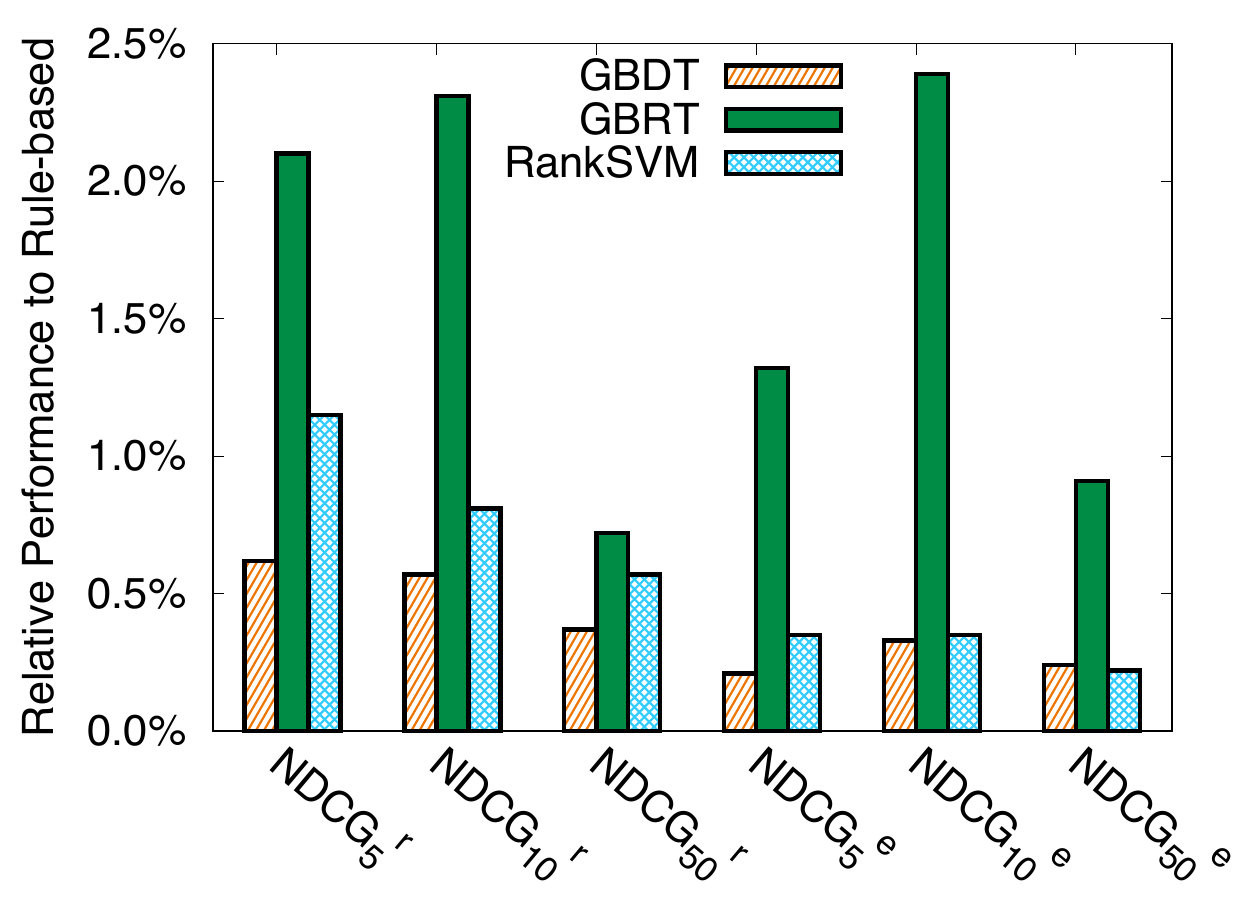}}
\subfigure[Online performance]{\includegraphics[width=0.495\columnwidth]{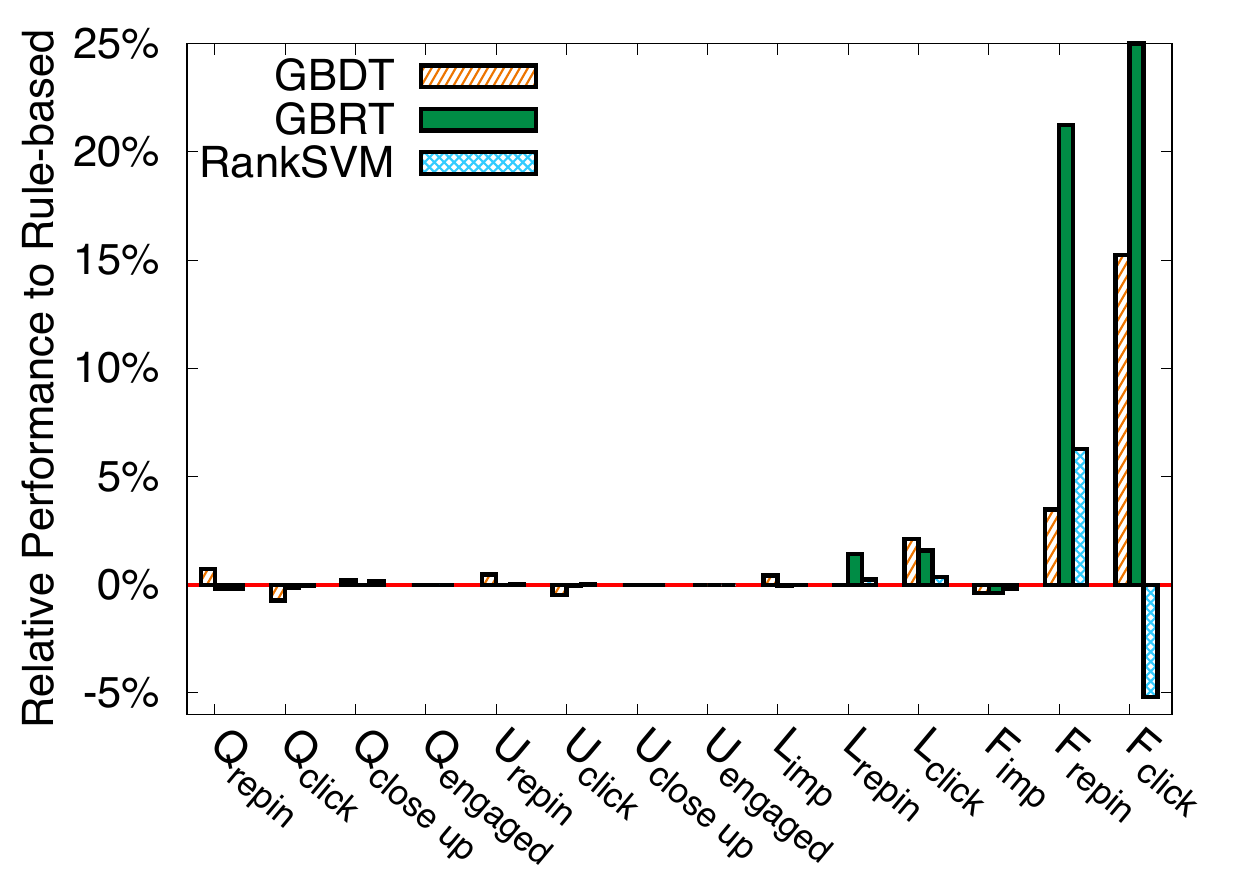}}
\caption{Relative performance of different models to the baseline Rule-based method in re-ranking stage.}\label{fig:results-rerank}
\vspace{-0.25cm}
\end{figure}

We spent much effort in feature engineering and migrate the rule-based ranking into machine-learned ranking. With multiple iterations of experiments, as shown in Figure~\ref{fig:results-rerank}, we are able to obtain comparable query-level and user-level performance with the rule-based methods and significantly outperformed the rule-based methods in terms of freshness and localness metrics. The click-through rate and repin rate on fresh pins is increased by 20\% when replacing the rule-based re-ranker with the \texttt{GBRT} model.

\section{Related Works}\label{sec-related}
Over the past decades, various ranking methods~\cite{BurgesRankNet2005,burges2010ranknet,cao2007learning,chapelle2011yahoo,dehghani2017neural,joachims2002optimizing,geng2007feature,yin2016ranking,liu2009learning,zheng2008general} have been proposed to improving the search relevance of web pages and/or user engagement in traditional search engine and e-commerce search engine. When we refer users to several tutorials~\cite{burges2010ranknet,liu2009learning} for more detailed introduction regarding the area of learning to rank, we focus on introducing how the applications of learning to rank for image search engine in industry evolves over time.

Prasad et. al.~\cite{prasad1987microcomputer} developed the first microcomputer-based image database retrieval system. After the successful launch of the Google Image Search Product in 2001, various image retrieval systems are deployed for public usage. Earlier works on image retrieval systems~\cite{datta2008image} focus on candidate retrieval with the image indexing techniques. 

In recent years, many works have been proposed to improve the ranking of the image search results using visual features and personalized features. For instance, Jing et al.~\cite{JingPAMI08} proposed the visualrank algorithm which ranks the Google image search results based on their centrality in visual similarity graph. On another hand, How to leverage user feedbacks and personalized signals for image ranking were studied in both Yahoo Image Corpora~\cite{OHareYahoo}, Flickr Image Corpora~\cite{fan2009justclick} and Pinterest Image Corpora~\cite{lo2016understanding}. In parallel to industry applications, research about Bayesian personalized ranking~\cite{RendleUAI} has been studied to improve the image search from implicit user feedbacks.

In addition to general image search products, recently many applications have also focused on specific domains such as fashion\footnote{https://www.shopstyle.com/}, food\footnote{www.supercook.com/}, home decoration\footnote{https://www.houzz.com/} etc. This trend also motivates researchers to focus on domain-specific image retrieval systems~\cite{HeAAAI2016,aizawa2015foodlog,liu2016deepfashion}. In Pinterest, while we have focused on the four verticals: fashion, food, beauty and home decoration, we also aim to help people discover the things they love for any domain.

\section{Conclusion and Future Works}\label{sec:con}
We introduced how we leverage user feedback into both training data and featurization to improve our cascading core ranking for Pinterest Image Search Engine. We empirically and theoretically analyzed various ranking models to understand how each of them performs in our image search engine. We hope those practical lessons learned from our ranking module design and deployment could also benefit other image search engines.

In the future, we plan to focus on two directions. First, as we have already observed good performance of both \texttt{DNN} and \texttt{CNN} ranking models, we plan to launch and serve them on-line directly instead of feeding their predicted scores as new features into tree-based ranking models. Second, many of our embedding-based features such as word embedding, visual embedding and user embedding were trained and shared across all the products in Pinterest such as home feed recommendation, advertisement, shopping etc. We plan to obtain the search-specific embedding features to understand the ``intents'' under the search scenario.

\vspace{0.1cm}
\noindent\textbf{Acknowledgement.} Thanks very much to the entire teams of engineers in search feature, search quality and search infra, especially to Chao Tan, Randall Keller, Wenchang Hu, Matthew Fong, Laksh Bhasin, Ying Huang, Zheng Liu, Charlie Luo, Zhongxian Cheng, Xiaofang Cheng, Xin Liu, Yunsong Guo for their much efforts and help in launching the ranking pipeline. Thanks to Jure Leskovec for valuable discussions.
\bibliographystyle{ACM}
\bibliography{ranking}


\begin{thebibliography}{37}


\ifx \showCODEN    \undefined \def \showCODEN     #1{\unskip}     \fi
\ifx \showDOI      \undefined \def \showDOI       #1{#1}\fi
\ifx \showISBNx    \undefined \def \showISBNx     #1{\unskip}     \fi
\ifx \showISBNxiii \undefined \def \showISBNxiii  #1{\unskip}     \fi
\ifx \showISSN     \undefined \def \showISSN      #1{\unskip}     \fi
\ifx \showLCCN     \undefined \def \showLCCN      #1{\unskip}     \fi
\ifx \shownote     \undefined \def \shownote      #1{#1}          \fi
\ifx \showarticletitle \undefined \def \showarticletitle #1{#1}   \fi
\ifx \showURL      \undefined \def \showURL       {\relax}        \fi
\providecommand\bibfield[2]{#2}
\providecommand\bibinfo[2]{#2}
\providecommand\natexlab[1]{#1}
\providecommand\showeprint[2][]{arXiv:#2}

\bibitem[\protect\citeauthoryear{Aizawa and Ogawa}{Aizawa and Ogawa}{2015}]%
        {aizawa2015foodlog}
\bibfield{author}{\bibinfo{person}{Kiyoharu Aizawa} {and}
  \bibinfo{person}{Makoto Ogawa}.} \bibinfo{year}{2015}\natexlab{}.
\newblock \showarticletitle{Foodlog: Multimedia tool for healthcare
  applications}.
\newblock \bibinfo{journal}{{\em IEEE MultiMedia\/}} \bibinfo{volume}{22},
  \bibinfo{number}{2} (\bibinfo{year}{2015}), \bibinfo{pages}{4--8}.
\newblock


\bibitem[\protect\citeauthoryear{Blei, Ng, and Jordan}{Blei
  et~al\mbox{.}}{2003}]%
        {blei2003latent}
\bibfield{author}{\bibinfo{person}{David~M Blei}, \bibinfo{person}{Andrew~Y
  Ng}, {and} \bibinfo{person}{Michael~I Jordan}.}
  \bibinfo{year}{2003}\natexlab{}.
\newblock \showarticletitle{Latent dirichlet allocation}.
\newblock \bibinfo{journal}{{\em Journal of machine Learning research\/}}
  \bibinfo{volume}{3}, \bibinfo{number}{Jan} (\bibinfo{year}{2003}),
  \bibinfo{pages}{993--1022}.
\newblock


\bibitem[\protect\citeauthoryear{Burges, Shaked, Renshaw, Lazier, Deeds,
  Hamilton, and Hullender}{Burges et~al\mbox{.}}{2005}]%
        {BurgesRankNet2005}
\bibfield{author}{\bibinfo{person}{Chris Burges}, \bibinfo{person}{Tal Shaked},
  \bibinfo{person}{Erin Renshaw}, \bibinfo{person}{Ari Lazier},
  \bibinfo{person}{Matt Deeds}, \bibinfo{person}{Nicole Hamilton}, {and}
  \bibinfo{person}{Greg Hullender}.} \bibinfo{year}{2005}\natexlab{}.
\newblock \showarticletitle{Learning to Rank Using Gradient Descent}. In
  \bibinfo{booktitle}{{\em Proceedings of the $22^{\texttt{nd}}$ International
  Conference on Machine Learning}}. \bibinfo{pages}{89--96}.
\newblock


\bibitem[\protect\citeauthoryear{Burges}{Burges}{2010}]%
        {burges2010ranknet}
\bibfield{author}{\bibinfo{person}{Christopher~JC Burges}.}
  \bibinfo{year}{2010}\natexlab{}.
\newblock \showarticletitle{From ranknet to lambdarank to lambdamart: An
  overview}.
\newblock \bibinfo{journal}{{\em Learning\/}} \bibinfo{volume}{11},
  \bibinfo{number}{23-581} (\bibinfo{year}{2010}).
\newblock


\bibitem[\protect\citeauthoryear{Cao, Qin, Liu, Tsai, and Li}{Cao
  et~al\mbox{.}}{2007}]%
        {cao2007learning}
\bibfield{author}{\bibinfo{person}{Zhe Cao}, \bibinfo{person}{Tao Qin},
  \bibinfo{person}{Tie-Yan Liu}, \bibinfo{person}{Ming-Feng Tsai}, {and}
  \bibinfo{person}{Hang Li}.} \bibinfo{year}{2007}\natexlab{}.
\newblock \showarticletitle{Learning to rank: from pairwise approach to
  listwise approach}. In \bibinfo{booktitle}{{\em Proceedings of the 24th
  international conference on Machine learning}}. ACM,
  \bibinfo{pages}{129--136}.
\newblock


\bibitem[\protect\citeauthoryear{Chapelle and Chang}{Chapelle and
  Chang}{2011}]%
        {chapelle2011yahoo}
\bibfield{author}{\bibinfo{person}{Olivier Chapelle} {and} \bibinfo{person}{Yi
  Chang}.} \bibinfo{year}{2011}\natexlab{}.
\newblock \showarticletitle{Yahoo! learning to rank challenge overview}.
\newblock  (\bibinfo{year}{2011}), \bibinfo{pages}{1--24}.
\newblock


\bibitem[\protect\citeauthoryear{Datta, Joshi, Li, and Wang}{Datta
  et~al\mbox{.}}{2008}]%
        {datta2008image}
\bibfield{author}{\bibinfo{person}{Ritendra Datta}, \bibinfo{person}{Dhiraj
  Joshi}, \bibinfo{person}{Jia Li}, {and} \bibinfo{person}{James~Z Wang}.}
  \bibinfo{year}{2008}\natexlab{}.
\newblock \showarticletitle{Image retrieval: Ideas, influences, and trends of
  the new age}.
\newblock \bibinfo{journal}{{\it Comput. Surveys}} \bibinfo{volume}{40},
  \bibinfo{number}{2} (\bibinfo{year}{2008}), \bibinfo{pages}{5}.
\newblock


\bibitem[\protect\citeauthoryear{Dehghani, Zamani, Severyn, Kamps, and
  Croft}{Dehghani et~al\mbox{.}}{2017}]%
        {dehghani2017neural}
\bibfield{author}{\bibinfo{person}{Mostafa Dehghani}, \bibinfo{person}{Hamed
  Zamani}, \bibinfo{person}{Aliaksei Severyn}, \bibinfo{person}{Jaap Kamps},
  {and} \bibinfo{person}{W~Bruce Croft}.} \bibinfo{year}{2017}\natexlab{}.
\newblock \showarticletitle{Neural Ranking Models with Weak Supervision}.
\newblock \bibinfo{journal}{{\em arXiv preprint arXiv:1704.08803\/}}
  (\bibinfo{year}{2017}).
\newblock


\bibitem[\protect\citeauthoryear{Dietterich}{Dietterich}{2000}]%
        {dietterich2000ensemble}
\bibfield{author}{\bibinfo{person}{Thomas~G Dietterich}.}
  \bibinfo{year}{2000}\natexlab{}.
\newblock \showarticletitle{Ensemble methods in machine learning}. In
  \bibinfo{booktitle}{{\em International workshop on multiple classifier
  systems}}. Springer, \bibinfo{pages}{1--15}.
\newblock


\bibitem[\protect\citeauthoryear{Fan, Keim, Gao, Luo, and Li}{Fan
  et~al\mbox{.}}{2009}]%
        {fan2009justclick}
\bibfield{author}{\bibinfo{person}{Jianping Fan}, \bibinfo{person}{Daniel~A
  Keim}, \bibinfo{person}{Yuli Gao}, \bibinfo{person}{Hangzai Luo}, {and}
  \bibinfo{person}{Zongmin Li}.} \bibinfo{year}{2009}\natexlab{}.
\newblock \showarticletitle{JustClick: Personalized image recommendation via
  exploratory search from large-scale Flickr images}.
\newblock \bibinfo{journal}{{\em IEEE Transactions on Circuits and Systems for
  Video Technology\/}} \bibinfo{volume}{19}, \bibinfo{number}{2}
  (\bibinfo{year}{2009}), \bibinfo{pages}{273--288}.
\newblock


\bibitem[\protect\citeauthoryear{Friedman}{Friedman}{2001}]%
        {friedman2001greedy}
\bibfield{author}{\bibinfo{person}{Jerome~H Friedman}.}
  \bibinfo{year}{2001}\natexlab{}.
\newblock \showarticletitle{Greedy function approximation: a gradient boosting
  machine}.
\newblock \bibinfo{journal}{{\em Annals of statistics\/}}
  (\bibinfo{year}{2001}), \bibinfo{pages}{1189--1232}.
\newblock


\bibitem[\protect\citeauthoryear{Geng, Liu, Qin, and Li}{Geng
  et~al\mbox{.}}{2007}]%
        {geng2007feature}
\bibfield{author}{\bibinfo{person}{Xiubo Geng}, \bibinfo{person}{Tie-Yan Liu},
  \bibinfo{person}{Tao Qin}, {and} \bibinfo{person}{Hang Li}.}
  \bibinfo{year}{2007}\natexlab{}.
\newblock \showarticletitle{Feature selection for ranking}. In
  \bibinfo{booktitle}{{\em Proceedings of the 30th annual international ACM
  SIGIR conference on Research and development in information retrieval}}. ACM,
  \bibinfo{pages}{407--414}.
\newblock


\bibitem[\protect\citeauthoryear{He and McAuley}{He and McAuley}{2016}]%
        {HeAAAI2016}
\bibfield{author}{\bibinfo{person}{Ruining He} {and} \bibinfo{person}{Julian
  McAuley}.} \bibinfo{year}{2016}\natexlab{}.
\newblock \showarticletitle{VBPR: Visual Bayesian Personalized Ranking from
  Implicit Feedback}. In \bibinfo{booktitle}{{\em Proceedings of the Thirtieth
  AAAI Conference on Artificial Intelligence}}. \bibinfo{pages}{144--150}.
\newblock


\bibitem[\protect\citeauthoryear{J\"{a}rvelin and
  Kek\"{a}l\"{a}inen}{J\"{a}rvelin and Kek\"{a}l\"{a}inen}{2002}]%
        {JarvelinNDCG2002}
\bibfield{author}{\bibinfo{person}{Kalervo J\"{a}rvelin} {and}
  \bibinfo{person}{Jaana Kek\"{a}l\"{a}inen}.} \bibinfo{year}{2002}\natexlab{}.
\newblock \showarticletitle{Cumulated Gain-based Evaluation of IR Techniques}.
\newblock \bibinfo{journal}{{\em ACM Trans. Inf. Syst.\/}}
  \bibinfo{volume}{20}, \bibinfo{number}{4} (\bibinfo{year}{2002}),
  \bibinfo{pages}{422--446}.
\newblock


\bibitem[\protect\citeauthoryear{Jing and Baluja}{Jing and Baluja}{2008}]%
        {JingPAMI08}
\bibfield{author}{\bibinfo{person}{Yushi Jing} {and} \bibinfo{person}{Shumeet
  Baluja}.} \bibinfo{year}{2008}\natexlab{}.
\newblock \showarticletitle{VisualRank: Applying PageRank to Large-Scale Image
  Search}.
\newblock \bibinfo{journal}{{\em IEEE Transactions on Pattern Analysis and
  Machine Intelligence\/}}  \bibinfo{volume}{30} (\bibinfo{year}{2008}),
  \bibinfo{pages}{1877--1890}.
\newblock


\bibitem[\protect\citeauthoryear{Jing, Liu, Kislyuk, Zhai, Xu, Donahue, and
  Tavel}{Jing et~al\mbox{.}}{2015}]%
        {jing2015visual}
\bibfield{author}{\bibinfo{person}{Yushi Jing}, \bibinfo{person}{David Liu},
  \bibinfo{person}{Dmitry Kislyuk}, \bibinfo{person}{Andrew Zhai},
  \bibinfo{person}{Jiajing Xu}, \bibinfo{person}{Jeff Donahue}, {and}
  \bibinfo{person}{Sarah Tavel}.} \bibinfo{year}{2015}\natexlab{}.
\newblock \showarticletitle{Visual search at pinterest}. In
  \bibinfo{booktitle}{{\em Proceedings of the 21th ACM SIGKDD International
  Conference on Knowledge Discovery and Data Mining}}. ACM,
  \bibinfo{pages}{1889--1898}.
\newblock


\bibitem[\protect\citeauthoryear{Joachims}{Joachims}{2002}]%
        {joachims2002optimizing}
\bibfield{author}{\bibinfo{person}{Thorsten Joachims}.}
  \bibinfo{year}{2002}\natexlab{}.
\newblock \showarticletitle{Optimizing search engines using clickthrough data}.
  In \bibinfo{booktitle}{{\em Proceedings of the eighth ACM SIGKDD
  international conference on Knowledge discovery and data mining}}. ACM,
  \bibinfo{pages}{133--142}.
\newblock


\bibitem[\protect\citeauthoryear{Li, Wu, and Burges}{Li et~al\mbox{.}}{2008}]%
        {li2008mcrank}
\bibfield{author}{\bibinfo{person}{Ping Li}, \bibinfo{person}{Qiang Wu}, {and}
  \bibinfo{person}{Christopher~J Burges}.} \bibinfo{year}{2008}\natexlab{}.
\newblock \showarticletitle{Mcrank: Learning to rank using multiple
  classification and gradient boosting}. In \bibinfo{booktitle}{{\em Advances
  in neural information processing systems}}. \bibinfo{pages}{897--904}.
\newblock


\bibitem[\protect\citeauthoryear{Liu, Rogers, Shiau, Kislyuk, Ma, Zhong, Liu,
  and Jing}{Liu et~al\mbox{.}}{2017a}]%
        {liu2017related}
\bibfield{author}{\bibinfo{person}{David~C Liu}, \bibinfo{person}{Stephanie
  Rogers}, \bibinfo{person}{Raymond Shiau}, \bibinfo{person}{Dmitry Kislyuk},
  \bibinfo{person}{Kevin~C Ma}, \bibinfo{person}{Zhigang Zhong},
  \bibinfo{person}{Jenny Liu}, {and} \bibinfo{person}{Yushi Jing}.}
  \bibinfo{year}{2017}\natexlab{a}.
\newblock \showarticletitle{Related Pins at Pinterest: The Evolution of a
  Real-World Recommender System}. In \bibinfo{booktitle}{{\em Proceedings of
  the 26th International Conference on World Wide Web Companion}}.
  \bibinfo{pages}{583--592}.
\newblock


\bibitem[\protect\citeauthoryear{Liu, Xiao, Ou, and Si}{Liu
  et~al\mbox{.}}{2017b}]%
        {Liu:2017:CRO:3097983.3098011}
\bibfield{author}{\bibinfo{person}{Shichen Liu}, \bibinfo{person}{Fei Xiao},
  \bibinfo{person}{Wenwu Ou}, {and} \bibinfo{person}{Luo Si}.}
  \bibinfo{year}{2017}\natexlab{b}.
\newblock \showarticletitle{Cascade Ranking for Operational E-commerce Search}.
  In \bibinfo{booktitle}{{\em Proceedings of the 23rd ACM SIGKDD International
  Conference on Knowledge Discovery and Data Mining}}.
  \bibinfo{publisher}{ACM}, \bibinfo{address}{New York, NY, USA},
  \bibinfo{pages}{1557--1565}.
\newblock


\bibitem[\protect\citeauthoryear{Liu et~al\mbox{.}}{Liu et~al\mbox{.}}{2009}]%
        {liu2009learning}
\bibfield{author}{\bibinfo{person}{Tie-Yan Liu} {et~al\mbox{.}}}
  \bibinfo{year}{2009}\natexlab{}.
\newblock \showarticletitle{Learning to rank for information retrieval}.
\newblock \bibinfo{journal}{{\em Foundations and Trends{\textregistered} in
  Information Retrieval\/}} \bibinfo{volume}{3}, \bibinfo{number}{3}
  (\bibinfo{year}{2009}), \bibinfo{pages}{225--331}.
\newblock


\bibitem[\protect\citeauthoryear{Liu, Luo, Qiu, Wang, and Tang}{Liu
  et~al\mbox{.}}{2016}]%
        {liu2016deepfashion}
\bibfield{author}{\bibinfo{person}{Ziwei Liu}, \bibinfo{person}{Ping Luo},
  \bibinfo{person}{Shi Qiu}, \bibinfo{person}{Xiaogang Wang}, {and}
  \bibinfo{person}{Xiaoou Tang}.} \bibinfo{year}{2016}\natexlab{}.
\newblock \showarticletitle{Deepfashion: Powering robust clothes recognition
  and retrieval with rich annotations}. In \bibinfo{booktitle}{{\em Proceedings
  of the IEEE Conference on Computer Vision and Pattern Recognition}}.
  \bibinfo{pages}{1096--1104}.
\newblock


\bibitem[\protect\citeauthoryear{Lo, Frankowski, and Leskovec}{Lo
  et~al\mbox{.}}{2016}]%
        {lo2016understanding}
\bibfield{author}{\bibinfo{person}{Caroline Lo}, \bibinfo{person}{Dan
  Frankowski}, {and} \bibinfo{person}{Jure Leskovec}.}
  \bibinfo{year}{2016}\natexlab{}.
\newblock \showarticletitle{Understanding behaviors that lead to purchasing: A
  case study of pinterest}. In \bibinfo{booktitle}{{\em Proceedings of the 22nd
  ACM SIGKDD International Conference on Knowledge Discovery and Data Mining}}.
  ACM, \bibinfo{pages}{531--540}.
\newblock


\bibitem[\protect\citeauthoryear{Mao, Xu, Jing, and Yuille}{Mao
  et~al\mbox{.}}{2016}]%
        {mao2016training}
\bibfield{author}{\bibinfo{person}{Junhua Mao}, \bibinfo{person}{Jiajing Xu},
  \bibinfo{person}{Kevin Jing}, {and} \bibinfo{person}{Alan~L Yuille}.}
  \bibinfo{year}{2016}\natexlab{}.
\newblock \showarticletitle{Training and evaluating multimodal word embeddings
  with large-scale web annotated images}. In \bibinfo{booktitle}{{\em Advances
  in Neural Information Processing Systems}}. \bibinfo{pages}{442--450}.
\newblock


\bibitem[\protect\citeauthoryear{Matveeva, Burges, Burkard, Laucius, and
  Wong}{Matveeva et~al\mbox{.}}{2006}]%
        {matveeva2006high}
\bibfield{author}{\bibinfo{person}{Irina Matveeva}, \bibinfo{person}{Chris
  Burges}, \bibinfo{person}{Timo Burkard}, \bibinfo{person}{Andy Laucius},
  {and} \bibinfo{person}{Leon Wong}.} \bibinfo{year}{2006}\natexlab{}.
\newblock \showarticletitle{High accuracy retrieval with multiple nested
  ranker}. In \bibinfo{booktitle}{{\em Proceedings of the 29th annual
  international ACM SIGIR conference on Research and development in information
  retrieval}}. ACM, \bibinfo{pages}{437--444}.
\newblock


\bibitem[\protect\citeauthoryear{McCandless, Hatcher, and
  Gospodnetic}{McCandless et~al\mbox{.}}{2010}]%
        {McCandlessLucene2010}
\bibfield{author}{\bibinfo{person}{Michael McCandless}, \bibinfo{person}{Erik
  Hatcher}, {and} \bibinfo{person}{Otis Gospodnetic}.}
  \bibinfo{year}{2010}\natexlab{}.
\newblock \bibinfo{booktitle}{{\em Lucene in Action, Second Edition: Covers
  Apache Lucene 3.0}}.
\newblock \bibinfo{publisher}{Manning Publications Co.},
  \bibinfo{address}{Greenwich, CT, USA}.
\newblock
\showISBNx{1933988177, 9781933988177}


\bibitem[\protect\citeauthoryear{O'Hare, de~Juan, Schifanella, He, Yin, and
  Chang}{O'Hare et~al\mbox{.}}{2016}]%
        {OHareYahoo}
\bibfield{author}{\bibinfo{person}{Neil O'Hare}, \bibinfo{person}{Paloma de
  Juan}, \bibinfo{person}{Rossano Schifanella}, \bibinfo{person}{Yunlong He},
  \bibinfo{person}{Dawei Yin}, {and} \bibinfo{person}{Yi Chang}.}
  \bibinfo{year}{2016}\natexlab{}.
\newblock \showarticletitle{Leveraging User Interaction Signals for Web Image
  Search}. In \bibinfo{booktitle}{{\em Proceedings of the 39th International
  ACM SIGIR Conference on Research and Development in Information Retrieval}}.
  \bibinfo{pages}{559--568}.
\newblock


\bibitem[\protect\citeauthoryear{Prasad, Gupta, Toong, and Madnick}{Prasad
  et~al\mbox{.}}{1987}]%
        {prasad1987microcomputer}
\bibfield{author}{\bibinfo{person}{BE Prasad}, \bibinfo{person}{Amar Gupta},
  \bibinfo{person}{Hoo-Min~D Toong}, {and} \bibinfo{person}{Stuart~E Madnick}.}
  \bibinfo{year}{1987}\natexlab{}.
\newblock \showarticletitle{A microcomputer-based image database management
  system}.
\newblock \bibinfo{journal}{{\em IEEE Transactions on Industrial
  Electronics\/}} \bibinfo{number}{1} (\bibinfo{year}{1987}),
  \bibinfo{pages}{83--88}.
\newblock


\bibitem[\protect\citeauthoryear{Raykar, Krishnapuram, and Yu}{Raykar
  et~al\mbox{.}}{2010}]%
        {raykar2010designing}
\bibfield{author}{\bibinfo{person}{Vikas~C Raykar}, \bibinfo{person}{Balaji
  Krishnapuram}, {and} \bibinfo{person}{Shipeng Yu}.}
  \bibinfo{year}{2010}\natexlab{}.
\newblock \showarticletitle{Designing efficient cascaded classifiers: tradeoff
  between accuracy and cost}. In \bibinfo{booktitle}{{\em Proceedings of the
  16th ACM SIGKDD international conference on Knowledge discovery and data
  mining}}. ACM, \bibinfo{pages}{853--860}.
\newblock


\bibitem[\protect\citeauthoryear{Rendle, Freudenthaler, Gantner, and
  Schmidt-Thieme}{Rendle et~al\mbox{.}}{2009}]%
        {RendleUAI}
\bibfield{author}{\bibinfo{person}{Steffen Rendle}, \bibinfo{person}{Christoph
  Freudenthaler}, \bibinfo{person}{Zeno Gantner}, {and} \bibinfo{person}{Lars
  Schmidt-Thieme}.} \bibinfo{year}{2009}\natexlab{}.
\newblock \showarticletitle{BPR: Bayesian Personalized Ranking from Implicit
  Feedback}. In \bibinfo{booktitle}{{\em Proceedings of the Twenty-Fifth
  Conference on Uncertainty in Artificial Intelligence}}.
  \bibinfo{pages}{452--461}.
\newblock


\bibitem[\protect\citeauthoryear{Robertson and Zaragoza}{Robertson and
  Zaragoza}{2009}]%
        {Robertson:2009:PRF:1704809.1704810}
\bibfield{author}{\bibinfo{person}{Stephen Robertson} {and}
  \bibinfo{person}{Hugo Zaragoza}.} \bibinfo{year}{2009}\natexlab{}.
\newblock \showarticletitle{The Probabilistic Relevance Framework: BM25 and
  Beyond}.
\newblock \bibinfo{journal}{{\em Found. Trends Inf. Retr.\/}}
  \bibinfo{volume}{3}, \bibinfo{number}{4} (\bibinfo{year}{2009}).
\newblock


\bibitem[\protect\citeauthoryear{Smiley and Pugh}{Smiley and Pugh}{2011}]%
        {smiley2011apache}
\bibfield{author}{\bibinfo{person}{D. Smiley} {and} \bibinfo{person}{D.E.
  Pugh}.} \bibinfo{year}{2011}\natexlab{}.
\newblock \bibinfo{booktitle}{{\em Apache Solr 3 Enterprise Search Server}}.
\newblock \bibinfo{publisher}{Packt Publishing, Limited}.
\newblock
\showISBNx{9781849516075}


\bibitem[\protect\citeauthoryear{Song, Taylor, Wen, Hon, and Yu}{Song
  et~al\mbox{.}}{2008}]%
        {song2008viewing}
\bibfield{author}{\bibinfo{person}{Ruihua Song}, \bibinfo{person}{Michael~J
  Taylor}, \bibinfo{person}{Ji-Rong Wen}, \bibinfo{person}{Hsiao-Wuen Hon},
  {and} \bibinfo{person}{Yong Yu}.} \bibinfo{year}{2008}\natexlab{}.
\newblock \showarticletitle{Viewing term proximity from a different
  perspective}. In \bibinfo{booktitle}{{\em European Conference on Information
  Retrieval}}. Springer, \bibinfo{pages}{346--357}.
\newblock


\bibitem[\protect\citeauthoryear{Yin, Hu, Tang, Daly, Zhou, Ouyang, Chen, Kang,
  Deng, Nobata, et~al\mbox{.}}{Yin et~al\mbox{.}}{2016}]%
        {yin2016ranking}
\bibfield{author}{\bibinfo{person}{Dawei Yin}, \bibinfo{person}{Yuening Hu},
  \bibinfo{person}{Jiliang Tang}, \bibinfo{person}{Tim Daly},
  \bibinfo{person}{Mianwei Zhou}, \bibinfo{person}{Hua Ouyang},
  \bibinfo{person}{Jianhui Chen}, \bibinfo{person}{Changsung Kang},
  \bibinfo{person}{Hongbo Deng}, \bibinfo{person}{Chikashi Nobata},
  {et~al\mbox{.}}} \bibinfo{year}{2016}\natexlab{}.
\newblock \showarticletitle{Ranking relevance in yahoo search}. In
  \bibinfo{booktitle}{{\em Proceedings of the 22nd ACM SIGKDD International
  Conference on Knowledge Discovery and Data Mining}}. ACM,
  \bibinfo{pages}{323--332}.
\newblock


\bibitem[\protect\citeauthoryear{Zhang}{Zhang}{2004}]%
        {zhang2004solving}
\bibfield{author}{\bibinfo{person}{Tong Zhang}.}
  \bibinfo{year}{2004}\natexlab{}.
\newblock \showarticletitle{Solving large scale linear prediction problems
  using stochastic gradient descent algorithms}. In \bibinfo{booktitle}{{\em
  Proceedings of the twenty-first international conference on Machine
  learning}}. ACM, \bibinfo{pages}{116}.
\newblock


\bibitem[\protect\citeauthoryear{Zheng, Chen, Sun, and Zha}{Zheng
  et~al\mbox{.}}{2007}]%
        {zheng2007regression}
\bibfield{author}{\bibinfo{person}{Zhaohui Zheng}, \bibinfo{person}{Keke Chen},
  \bibinfo{person}{Gordon Sun}, {and} \bibinfo{person}{Hongyuan Zha}.}
  \bibinfo{year}{2007}\natexlab{}.
\newblock \showarticletitle{A regression framework for learning ranking
  functions using relative relevance judgments}. In \bibinfo{booktitle}{{\em
  Proceedings of the 30th annual international ACM SIGIR conference on Research
  and development in information retrieval}}. ACM, \bibinfo{pages}{287--294}.
\newblock


\bibitem[\protect\citeauthoryear{Zheng, Zha, Zhang, Chapelle, Chen, and
  Sun}{Zheng et~al\mbox{.}}{2008}]%
        {zheng2008general}
\bibfield{author}{\bibinfo{person}{Zhaohui Zheng}, \bibinfo{person}{Hongyuan
  Zha}, \bibinfo{person}{Tong Zhang}, \bibinfo{person}{Olivier Chapelle},
  \bibinfo{person}{Keke Chen}, {and} \bibinfo{person}{Gordon Sun}.}
  \bibinfo{year}{2008}\natexlab{}.
\newblock \showarticletitle{A general boosting method and its application to
  learning ranking functions for web search}. In \bibinfo{booktitle}{{\em
  Advances in neural information processing systems}}.
  \bibinfo{pages}{1697--1704}.
\newblock


\end{thebibliography}
\end{document}